\documentclass[11pt,a4paper]{article}

\usepackage[utf8]{inputenc} 
\usepackage[T1]{fontenc}    %
\usepackage[english]{babel} 
\usepackage[final]{microtype} 

\usepackage{amssymb,amsmath,mathtools} 
\usepackage{amsthm} 

\usepackage{xcolor}
\usepackage{bbm}
\usepackage{braket}

\usepackage[hmargin=0.12\paperwidth,vmargin=0.16\paperwidth,bindingoffset=0cm]%
{geometry} 

\numberwithin{equation}{section} 

\theoremstyle{plain}
  
  \newtheorem{prop}{Proposition}
  
  \newtheorem{cor}{Corollary}
\theoremstyle{definition}
  
  \newtheorem{remark}{Remark}
  
  \numberwithin{prop}{section}
   \numberwithin{cor}{section}
   \numberwithin{remark}{section}

\newcommand{\MeijerG}[8][\Big]{G^{{ #2 },{ #3 }}_{{ #4 },{ #5 }} #1( \begin{matrix} #6 \\ #7 \end{matrix}\, #1\vert\, #8 #1)}

\title{\Large\bfseries The random matrix hard edge --- applications, exact PDFs and large $N$ expansions}

\author{Peter J. Forrester${}^1$,
Anas A. Rahman${}^2$, and 
Bo-Jian Shen${}^1$}
\date{}
\date{}

\begin{document}

\maketitle

${}^1$ School of Mathematics and Statistics,  The University of Melbourne,
Victoria 3010, Australia. \: \: Email: {\tt pjforr@unimelb.edu.au};  {\tt bojian.shen@unimelb.edu.au}\\

${}^2$
Department of Mathematics,
The University of Hong Kong, Hong Kong. \\
Email: {\tt aarahman@hku.hk}

\date{}

\begin{abstract}
A hard edge of a Hermitian random matrix occurs when the  eigenvalue density is strictly zero on one side of the edge, and nonzero on the other. Another possibility is that the eigenvalues occur in $\pm$ pairs, which creates a hard edge at the origin for the positive eigenvalues. Both review material 
and new results are presented for random matrix ensembles exhibiting a hard edge. The review material has two focuses: one is on applications of the random matrix hard edge, while the other is on exact eigenvalue probability density functions.
The new results relate to the large $N$ expansion of correlation functions and conditioned gap probability generating functions with hard edge scaling for the Jacobi ensemble. For all even $\beta$ we are able to show that these quantities admit expansions in inverse even powers of the variable $N_{\rm hJ}':= N - 1/2 + (a + b + 1)/\beta$. For the $\beta =2$ Jacobi unitary ensemble, we can relate the leading (next leading) correction to the limit law to a particular second order 
(fourth order) differential operator applied to the latter. The analogous second order differential operator is found for $\beta=1,4$, leading then to its conjectured form for all $\beta>0$.
\end{abstract}

\maketitle

\section{Introduction}
The term ``hard edge'' in random matrix theory was coined in the 1994 paper \cite{TW94b} in relation to the neighbourhood of the origin of 
the eigenvalues of complex Wishart matrices $W = X^\dagger X$. Here $X$ is an $n \times N$ ($n \ge N$) standard complex Gaussian matrix. The matrices $W$ being positive definite constrains the eigenvalues to be positive. Furthermore, for $N$ large and $n-N$ fixed the eigenvalue support is to leading order $(0,4N)$ \cite[Prop.~3.2.3]{Fo10}.
Thus the origin is a spectral edge, and furthermore
the density of eigenvalues on the negative real axis is strictly zero, explaining the adjective ``hard''. Known in the literature prior to \cite{TW94b} was that the neighbourhood of the hard edge for complex Wishart matrices admits a well defined limiting statistical state upon the scaling of the eigenvalues, $\lambda_j \mapsto \lambda_j/(4N)$, ($j=1,2,\dots,N$), with $0 < \lambda_1 < \lambda_2 < \cdots< \lambda_N$ and taking $N \to \infty$ \cite{Ed88,Fo93}. 

In this work, we present some review material relating to the hard edge (\S \ref{S2} and \S \ref{S3}), as well as some original research.
The review material of \S \ref{S2} is on applied settings which give rise to the need for analytic results relating to particular hard edge scaled states. 
The first of these --- condition number of a large linear system summarised in \S \ref{S2.1} --- is the setting of what appears to be the first consideration of random matrix hard edge scaling in the
literature. This was in the field of numerical analysis, and dates from 1988 \cite{Ed88}. The relevance of the hard edge scaled state to problems in quantum mechanics followed in the early 1990's. This involved settings in both relativistic quantum field theory (\S \ref{S2.2}) and mesoscopic quantum transport (\S \ref{S2.3}). Later in the 1990's applications of the random matrix hard edge to topological superconductors were found (\S \ref{S2.4}), which over a decade later linked with studies on Majorana fermions. In the early 2000's applications in number theory (\S \ref{S2.5}) and asymptotic combinatorics (\S \ref{S2.6}) appeared.

The review material of \S \ref{S3} is on specifying random matrix ensembles with a hard edge and admitting explicit joint eigenvalue (or singular value) probability density functions (PDFs).
The most precise analytic results relating to the hard edge often begin from knowledge of the 
joint eigenvalue PDFs for classes of positive definite matrices. 
The Wishart, chiral and Jacobi ensembles are the best known of this type (\S \ref{S3.1} and \S \ref{S3.2}). They also permit dynamical generalisations involving Brownian motion evolutions, and generalisations involving parametric sums (\S \ref{S3.3}). Product ensembles provide another rich source of explicit eigenvalue PDFs (\S \ref{S3.4}).

\section{Some applied settings}\label{S2}
\subsection{Condition number of a large linear system}\label{S2.1}
Here, following the classic works 
\cite{vNG47} and \cite{Sm85}, the question of interest is the stability of a large linear system under perturbation of the coefficients. Let there be $N$ equations and $N$ unknowns, and write the linear system in matrix form $A \mathbf x = \mathbf b$, where $A$ is the coefficient matrix, $\mathbf x$ is the column vector of unknowns and
$\mathbf b$ is the column vector of constants giving the right hand side of the equations when written in conventional form. Now perturb the coefficient matrix $A \mapsto A + \delta A$ to obtain the new linear system
$(A + \delta A) \mathbf y = \mathbf b$. 
To quantify the stability, one would like 
to relate $|| \mathbf y - \mathbf x||/|| \mathbf y|| $ to 
$\varepsilon := || \delta A||/ ||A||$. For this purpose the condition number $\kappa(A)$ is introduced by the requirement
\begin{equation}\label{1}
{|| \mathbf y - \mathbf x|| \over || \mathbf y||} \le
\varepsilon \kappa(A).
\end{equation}
Moreover, taking the matrix norm as the spectral norm, it is straightforward to show that this inequality is satisfied by choosing $\kappa(A) = \mu_N/\mu_1$, that is the ratio of the biggest to smallest singular values of $A$. Equivalently, $\kappa(A) = ( \lambda_N/ \lambda_1)^{1/2}$,
where $\lambda_1$ ($\lambda_N$) denotes the smallest
(largest) eigenvalue of the positive definite matrix $A^\dagger A$.

In the 1988 work \cite{Ed88}, a large $N$ analysis of $\kappa(A)$ for $A$ an $N \times N$ standard Gaussian matrix was undertaken. The matrix $W = A^T A$ is an example of a real Wishart matrix. As for the corresponding complex Wishart matrices, for large $N$ the leading order support of the spectrum is $(0,4N)$. Thus to leading order $\lambda_N$ is equal to $4N$. The determination of the condition number (which here is a random variable) then requires a determination of the statistical properties of the smallest eigenvalue, which is a hard edge property.

\subsection{Low energy eigenvalues of random Dirac operators in QCD}\label{S2.2}
Seminal works from the 1950's and early 1960's put forward the eigenvalues of large Hermitian random matrices as statistical models of highly excited energy states of heavy nuclei --- the 1965 reference \cite{Po65} reproduces many of the original works, with commentary. Essential here was the role played by a global time reversal symmetry $K$. Depending on whether $K^2=1$ or $K^2 = -1$, the Hermitian matrix is required to be real symmetric, or quaternion self dual respectively. In the mid 1980's it was realised that such an hypothesis is not restricted to complex quantum systems such as heavy nuclei, but rather applies equally well to simple billiard quantum systems, provided the underlying classical mechanics is chaotic \cite{BGS84}.

In relativistic quantum mechanics, it is a celebrated result that the eigenvalues of ($i$ times) the massless Dirac operator come in symmetric pairs $\pm \lambda$ around zero, $\lambda > 0$. Using the chiral basis, in the early 1990s it was realised
\cite{Ve94} that the appropriate matrix structure to model this situation, using an eigenbasis of $\gamma_5$, is
\begin{equation}\label{2}
\begin{bmatrix} 0_{n \times n} & X_{n \times N} \\
(X^\dagger)_{N \times n} & 0_{N \times N}
\end{bmatrix}.
\end{equation}
With $n \ge N$, here there are $n-N$ zero eigenvalues, while in general the remaining $2N$ eigenvalues come in pairs $\pm \mu_i$ ($i=1,\dots,N$), where $\{ \mu_i \}$ are the non-zero singular values of $X$, or equivalently the positive square roots of the eigenvalues of $X^\dagger X$. Moreover, analogous to the time reversal symmetry constraints imposed on Hermitian random matrices modelling a non-relativistic Hamiltonian, it was realised that the symmetries of a gauge field
 included as an additive term to the free Dirac operator for QCD applications
lead to real symmetric and self-dual quaternion subcases.
Upon the specification of the appropriate symmetry class, it is now the low-lying Dirac eigenvalues which are hypothesised to have statistical properties identical to those of  (\ref{2}) with $X$ therein a Gaussian random matrix. Thus it is the hard edge that is relevant, and comparisons are possible after the hard scaling limit has been carried out. Most accessible is the one point density \cite{DHK99}.

\subsection{Fluctuation of quantum conductance}\label{S2.3}
Work from the 1990's, reviewed in \cite{Be97}, relates quantum conductance problems in mesoscopic systems (here to be taken as a quantum dot) to the eigenvalues of sub-blocks of unitary random matrices. The setting is to have two leads, one with $N$ and the other with $n$ channels. The amplitudes associated with  the first are to be denoted
$\mathbf{I}$ ($\mathbf{O}$) for flow in (out), while the amplitudes associated with the second are to be denoted
$\mathbf{I} '$ ($\mathbf{O} '$) for flow in (out). These amplitudes are related by the $(N+n) \times (N+n)$ 
unitary scattering matrix $S$ according to
\begin{equation}\label{3}
S \begin{bmatrix} \mathbf{I} \\ \mathbf{I}' \end{bmatrix} =
\begin{bmatrix} \mathbf{O} \\ \mathbf{O}' \end{bmatrix}.
\end{equation}
As is consistent with (\ref{3}), the scattering matrix is decomposed into reflection $(R,R')$ and transmission $(T,T')$ submatrices according to
\begin{equation}\label{4}
S = \begin{bmatrix} R_{N \times N} & T_{N \times n}' \\
T_{n \times N} & R_{n \times n}' \end{bmatrix}.
\end{equation}
According to the Landauer formula, the dimensionless conductance $G/G_0$ ($G_0 = 2 e^2/ h$ is twice the fundamental quantum unit of conductance) is given by
$G/G_0 = {\rm Tr}(T^\dagger T) = {\rm Tr}((T')^\dagger T') $.
To proceed further, one hypothesises chaotic scattering within the quantum dot. With no global time reversal symmetry, $S$ is then well modelled by a Haar distributed unitary matrix. With time reversal symmetry $K$, from the foundational work early in the 1960's \cite{Dy62} it is known that $S$ is further constrained: $S = S^T$ ($K^2=1$),
$S = S^D$ ($K^2 = -1$), where the subscript $D$ denotes the quaternion dual. In all three cases the exact joint PDF of the eigenvalues $\{ t_j \}$ 
of
$T^\dagger T$ can be computed as examples of the Jacobi ensemble \cite{Be97,Fo06}; see also \S \ref{S3.2}, paragraph following \eqref{1.3e} below.

With $T$ being subunitary, one has $0 < t_j < 1$. Thus there are two hard edges, one at 0 and the other at 1. This number can be reduced to one, by the change of variables $t_j = 1/(1 + \lambda_j)$. This effectively isolates the hard edge at 1 in the original variables, where the typical eigenvalue spacing is of order $1/N^2$. In the new variables, the dimensionless conductance is the linear statistic
$a(\lambda_j)=\sum_{j=1}^N 1/(1 + \lambda_j)$. Thus the variation of this statistic relative to the typical eigenvalue spacing is very slow. Such a situation was already isolated in the pioneering work of 1963 \cite{DM63}, although in the bulk of the random matrix spectrum rather than at the edge. The method of analysis proceeded by first computing the scaling limit, whereby the average spacing between eigenvalues is of order unity. Then in the fluctuation formula for the variance (see e.g.~\cite[Prop.~2.1]{Fo23})
\begin{equation}\label{5}
{\rm Var}\Big ( \sum_j a(y_j) \Big ) =
- {1 \over 2} \int d y_1 \int d y_2 \,
(a(y_2) - a(y_1))^2 \rho_{(2)}^{T, {\rm s}}(y_1, y_2),
\end{equation}
where $\rho_{(2)}^{T, {\rm s}}(y_1, y_2)$ denotes the scaled truncated two-particle correlation. However, in this formula the fact that in the original variables $a(\lambda)$ is slowly varying is yet to be accounted for. This is done by the replacement $a(y) \mapsto a(y/\alpha)$, and then taking the limit $\alpha \to \infty$. A simple change of variables shows that this limit reduces to computing
\begin{equation}\label{6}
\lim_{\alpha \to \infty} \alpha^2 \rho_{(2)}^{T, {\rm s}}(\alpha y_1, \alpha y_2)
\end{equation}

\subsection{Topological superconductors and Majorana fermions}\label{S2.4}
In the study of properties around the Fermi level $E=0$ of topological superconductors, it was realised in 1997 \cite{AZ97} that the consideration of time reversal symmetry must be combined with that of particle hole symmetry. As for time reversal, the corresponding operator, $\mathcal C$ say,
is required to be anti-unitary and so square to plus or minus the identity. However, in distinction to time reversal, $\mathcal C$ is required to anti-commute with the Hamiltonian matrix,
$H = -\mathcal C H \mathcal C^{-1} $. This symmetry for $\mathcal C^2 = \mathbb I$
 is exhibited by the block matrix structure (each block $N \times N$, with $N$ even for spin degrees of freedom)
\begin{equation}\label{6a}
H = \begin{bmatrix}
H_0 & \Delta \\ - \bar{\Delta} & - H_0^T 
\end{bmatrix},
\end{equation}
where $H_0 = H_0^\dagger$ and $\Delta = - \Delta^T$, as appears in the matrix Bogoliubov-deGennes equation. This can be seen by writing
$\mathcal C = U \mathcal K$, with $U$ a
$2N \times 2 N$ unitary and $\mathcal K$ corresponding to complex conjugation (this is true in general), and making the choice
\begin{equation}\label{6a+}
U = \begin{bmatrix} 0_N & I_N \\
I_N & 0_N \end{bmatrix}.
\end{equation}

As for (\ref{2}), the eigenvalues of (\ref{6a}) come in $\pm$ pairs.  In combination with the consideration of time reversal symmetry, particle-hole symmetry gives rise to four additional matrix classes beyond those relating to (\ref{2}):
\begin{equation}\label{6b}
H = i A^{\rm r}, \quad H = i A^{\rm qr}, \quad H = \begin{bmatrix} 0_N & C \\
C^\dagger & 0_N \end{bmatrix} \: (C = C^T), \quad
H = \begin{bmatrix} 0_N & C \\
C^\dagger & 0_N \end{bmatrix} \: (C = C^D).
\end{equation}
Here $A^{\rm r}$ is an $N \times N$ real anti-symmetric matrix,
$A^{\rm qr}$ is an $N \times N$ ($N$ even) anti-symmetric matrix which when written in $N \times N$ block form has independent entries given by $2 \times 2$ matrices corresponding to quaternions with all entries real. The matrix $C$ in the third of these is required to be complex symmetric, while in the final matrix it is required to be quaternion self dual (in relation to this, see
\cite[Eq,~(1.19)]{Fo10}).

The Hamiltonian matrix $H$ relates to the Hamiltonian $\mathcal H$ of the superconductor (without spin degrees of freedom) via $\mathcal H = \psi^\dagger H \psi$ with 
$\psi = (c_1,\dots,c_N, c_1^\dagger,\dots, c_N^\dagger)$ having as its components Fermi operators. The change of basis from $\{ c_i, c_i^\dagger\}$ to real (Majorana fermion) operators
$ \gamma_{2j-1} = c_j + c_j^\dagger$, $
\gamma_{2j} =-i (c_j -
c_j^\dagger)$ effectively transforms $U$ in
(\ref{6a+}) to $I_{2N}$. The particle hole symmetry requirement then simplifies to $H = - \bar{H}$, which for $H$ Hermitian gives the first class in (\ref{6b}). We refer to 
\cite{Be15} for more on the Majorana fermion context of topological superconductors.

Gaussian random matrix ensembles come about by choosing the independent entries implied by the structures (\ref{6b}) to be standard Gaussians from the appropriate number field. As for the Dirac eigenvalues, of interest is the statistical properties of the lowest lying eigenvalues, and thus it is the hard edge that is again relevant.

\subsection{Non-intersecting lattice paths and tilings in a half space}
Consider the square lattice on the quadrant $x\ge 0, y \ge 0$. At  co-ordinates $(0,2i-2)$ $(i=1,\dots,p)$ mark in points of what are to be the beginning of $p$ non-intersecting lattice paths, and at $(2M, 2i-2)$ $(i=1,\dots,p)$ mark in points that are to be the endpoints. The lattice paths are formed by the rule that they consist of $2M$ segments, each one unit up and to the right, or one unit down and to the right. Each path is conditioned not to intersect its neighbour (starting on a sublattice with a definite parity guarantees that the paths cannot cross), with the bottom path furthermore conditioned not to go below $y=0$. Such a configuration of non-intersecting lattice paths is in bijection with rhombi tiling of the half hexagon; see \cite[Fig.~1]{FN09}. One notes that the neighbourhood of the bottom path is a 
hard edge.

After appropriate scaling, these non-intersecting lattice paths admit a continuum limit to $p$ non-intersecting Brownian bridges with a reflecting boundary condition at the origin
(what was the $x$-coordinate becomes time, and the $y$-direction space). For any fixed time, the positions of the $p$ Brownian particles have the same probability density function as that of $i$ times $2p \times 2p$ real anti-symmetric matrices \cite{KTNK03}. 

There is another relation between the rhombi tiling model and the eigenvalues of real anti-symmetric random matrices. This comes about by supplementing the positions of the corresponding lattice paths at each step by marking the positions of holes, these being lattice sites that were not chosen. At step 2 and 3 there is one such site, at steps 4 and 5 there are two such sites, etc.~see \cite[Fig.~1]{FN09}. A space like continuum limit can be taken in which these holes have the same joint distribution as that of the so-called anti-symmetric GUE minor process. In the latter, at each step, one row and one column is added to $i$ times a $k \times k$ real anti-symmetric matrix to form an $i$ times a $k+1 \times k+1$ real anti-symmetric matrix, with the quantity of interest being the joint eigenvalue probability density function. There is also an analogous interlaced particle system associated with domino tiling of the half Aztec diamond \cite{FF11}. 

A generalisation of the model of $p$ non-intersecting Brownian bridges with a reflecting boundary condition at the origin is that of the non-intersecting squared Bessel process in the presence of a wall 
\cite{De13}. This in turn relates to the eigenvalues of
random matrices of the form \eqref{2}, now with Brownian entries with conditioned initial and final values. An effect of interest is the appearance of a tacnode kernel, when the trajectories touch the origin for the first time; see
\cite[Fig.~2]{De13}. In the case of Brownian motion on orthogonal or unitary symplectic matrices starting from the origin, one encounters the symmetric Pearcey kernel when the eigenvalues first reach the hard edge at $\theta = \pi$
\cite{LW17}.

\subsection{Statistical properties of $L$-functions}\label{S2.5}
The Riemann zeta function, defined for Re$(s) > 1$ as $\zeta(s) =
\sum_{n=1}^\infty {1 \over n^s}$, is an arithmetic function due to its Euler product form $\zeta(s) = \prod_{p}(1 - p^{-s})^{-1}$, where the product is over all primes. Another key property is  a functional equation, relating $\zeta(s) $ to $\zeta(1-s)$, and thus with the line $s = 1/2 + it$, $t \in \mathbb R$ is fixed by the mapping. According to the Riemann hypothesis, all complex zeros of $\zeta(s)$ lie on this line.
A combination of analytic and numerical evidence has led to a conjecture, known as the Montgomery-Odlyzko law, which asserts
that for large $t$ these zeros — referred to as the Riemann zeros — have the same statistical properties as the bulk eigenvalues of large Hermitian complex Gaussian matrices; see \cite{Od87}. The 2001 work \cite{KS01} presented a refinement by introducing Haar distributed unitary matrices to quantify finite size effects in relation to the Riemann zeros. One recalls that in the large matrix size limit,  Hermitian complex Gaussian matrices in the bulk of the spectrum and Haar distributed unitary matrices have the same statistical properties; see e.g.~\cite[\S 7.1.1 and \S 7.2.7]{Fo10}.

Generalising the Riemann zeta function are the class of arithmetic functions known as $L$-functions. As for $\zeta(s)$, they can, for Re$(s)>1$, be written in the form $L(s) = \sum_{n=1}^\infty {a_n \over n^s}$. Further, they admit a generalisation of the Euler product, as well as a functional equation fixing the line $s = 1/2 + it$, with all complex zeros conjuctured to lie on this line. The 1999 work \cite{KS99} 
relates the statistical properties of the zeros on the critical line nearest the real axis for families of $L$-functions
(for example, all those with $a_n = \chi_d(n)$, where $\chi_d(n)$ is the Kronecker symbol, for all allowed $d$)
to the eigenvalues of large random matrices from the classical unitary groups with Haar measure near $\theta = 0$. In addition to the unitary groups, real orthogonal and symplectic unitary matrices are also of this type. 
Review material on this line can be found in
\cite{Co06}.
Most significant from the viewpoint of the hard edge is that the eigenvalues of real orthogonal and symplectic unitary matrices come in complex conjugate pairs. Thus $\theta =0$ (along with $\theta = \pi$) is a hard edge for the independent eigenvalues.

\subsection{Longest increasing subsequences and growth processes}\label{S2.6}
A permutation of the set of numbers $\{1,2,\dots,N\}$ is one of its $N!$ different orderings of these numbers. For example, with $N=6$, one of the permutations is $\sigma=431652$. Contained within a given permutation are subsequences that are strictly increasing, with the number of elements it contains referred to as its length. For example, one sees that $\sigma$ contains increasing subsequences of lengths 1 and 2 only. Let $\ell_N$ denote the maximum of these lengths. A question of long standing interest
\cite{Ha56} is the statistical properties of this length, when $\sigma$ is chosen uniformly at random and $N$ is large.
Denoting the number of permutations of $\{1,2,\dots,N\}$ with $\ell_N \le a$
by $T_a(N)$, one sees that
${\rm Pr}(\ell_N \le a) = T_a(N)/N!$. 

Suppose now that the number $N$ itself is a Poisson random variable, chosen with probability $e^{-z^2} z^{2N}/N!$. In 
a sample of the resulting ensemble of random permutations, let $\ell$ denote the random variable for the maximum length of all the increasing subsequences. We have
\begin{equation}\label{7}
{\rm Pr} ( \ell \le a;z) = e^{-z^2} \bigg ( 1 +
\sum_{N=1}^\infty {T_a(N) \over (N!)^2} z^{2 N} \bigg ).
\end{equation}
The 1998 result \cite{Ra98} expresses the infinite sum herein as a particular matrix integral over the
group of $a \times a$ unitary matrices with Haar measure
\begin{equation}\label{8}
{\rm Pr} ( \ell \le a;z) = e^{-z^2}
\Big \langle e^{z {\rm Tr}(U + U^\dagger)} 
\Big \rangle_{ U \in {\rm U}(a)}
= e^{-z^2}\det [ I_{j-k}(2z) ]_{j,k=1,\dots,a},
\end{equation}
where the second equality, in which $I_\nu(z)$ denotes the I-Bessel function, is an equivalent form which was in fact deduced earlier in \cite{Ge90}. Both the matrix integral form and its Toeplitz determinant equivalent were deduced in separate works from  1994 \cite{Fo93c,FH94} on the hard edge scaling of the probability that there are no eigenvalues in the interval $(0,s)$ --- a so called gap probability --- for complex Wishart matrices with $n-N = a$. Hard edge scaling replaces $s$ by $s/(4N)$ as in keeping with the specification in the first paragraph of the Introduction, and the limit $N \to \infty$ is taken with $a$ fixed. Now replacing $s$ by $4z^2$ gives precisely the right hand side of (\ref{8}). The resulting expression for ${\rm Pr} ( \ell \le a;z)$
was used in the 2003 work \cite{BF03} to rederive the 1999 limit law for $\ell_N$ \cite{BDJ98} via the hard-to-soft edge transition,
\begin{equation}\label{9}
\lim_{N \to \infty} {\rm Pr} \Big ( {\ell_N - 2 \sqrt{N} \over N^{1/6}} \le s \Big ) = E_2^{\rm soft}((s,\infty)),
\end{equation}
where $E_2^{\rm soft}((s,\infty))$ is the soft edge scaled probability of no eigenvalues in $(s,\infty)$ of the Gaussian unitary ensemble \cite{Fo93,TW94a}.

Closely related to the increasing subsequence problem are certain polynuclear growth models and last passage percolation problems; see e.g.~\cite[Ch.~10]{Fo10}. In these studies formulas arise for  quantities analogous to Pr$(\ell \le a;z)$
which also permit interpretations as random matrix gap probabilities at a hard edge, and which moreover permit a well defined hard edge scaled limit \cite{FR02b,BO25,BJS26}.

 \subsection{Hermitian and non-Hermitian hard edge  density duality}\label{S2.7}
 We have seen that several block off-diagonal matrix structures underlie Hermitian matrix ensembles with a hard edge, in particular \eqref{2} and \eqref{6b}. In the next section it will be revised that with independent (up to symmetry in the cases of \eqref{6b}) Gaussian entries from the specified number field (real, complex or quaternion) these ensembles permit exact determination of their eigenvalue PDFs. From this starting point, which shows that the eigenvalue distributions are examples of determinantal or Pfaffian point processes (see e.g.~\cite{Fo10}), one can proceed to calculate all the correlation functions and their hard edge scaled limits. In fact the general $k$-point correlation function is determined by a single correlation kernel $K_N(x,y)$, independent of $k$, which is a scalar in the determinantal cases and a $2 \times 2$ matrix in the Pfaffian cases. The correlation kernel in turn permits a well defined hard edge scaling limit.

 Consider now the non-Hermitian matrix $X_{n \times N}$ in \eqref{2}. In the case $n = N$ the three choices of number field give rise to what is termed the Ginibre ensembles 
 \cite{Gi65}. Of these, the complex case is distinguished by the eigenvalue PDF being rotationally invariant in the complex plane. A further class of non-Hermitian random matrices with this latter property is given by the block structure
 \begin{equation}\label{1.1v}
 \begin{bmatrix} 0_{n \times n} & X_{n \times N} \\
 Y_{N \times n} & 0_{N \times N} \end{bmatrix},
\end{equation} 
with $X,Y$ having independent standard complex Gaussian entries. This is known from symmetry consideration akin to 
\cite{AZ97}, now applied directly to the non-Hermitian setting
\cite{BL02,Ma07,HKKU20}, where complex conjugation and transpose are distinct operations $\bar{M} \ne M^T$. With $n \ge N$, as well as there being $n - N$ zero eigenvalues, the eigenvalues come in $\pm$
pairs of the square root of the eigenvalues of
the non-Hermitian product matrix $YX$. In distinction to the complex Ginibre ensemble, there is no translation invariance near the origin in the $N \to \infty$ limit. Instead the origin acts as a generalised hard edge.

It has been known since the 2004 work \cite{Os04} that the complex eigenvalues of $ 2 YX$ (the factor of 2 is for later convenience) form a determinantal point process. All the $k$-point correlation functions are readily computed. 
Using squared variables, $|z| \mapsto |z|^2$,
the $n,N \to \infty$ limit of the eigenvalue density is,
with $\nu = n - N$ and $I_\nu(x), K_\nu(x)$ particular Bessel functions in standard notation,
\begin{equation}\label{1.2v}
\pi R_{1,\nu}^{\rm NH}(z) = 
2 u I_\nu(u) K_\nu(u) \Big |_{u = |z|^2}.
\end{equation}
For $t > 0$, introduce now the transform
\begin{equation}\label{1.3v}
\mathcal P[f](t) = {1 \over \pi} 
\int_{-\infty}^\infty {t/(2 \pi) \over u^2 + (t/(2 \pi))^2} f(u) \, du,
\end{equation}
corresponding to the harmonic extension of $f$ from the real axis to the upper half complex plane. A very recent advance
\cite{CXLR26} uses \eqref{1.2v} to
link $\pi R_{1,\nu}^{\rm NH}(z)$ to the spectral density
$R_{1,\nu}^{\rm H}(x)$, $x > 0$, of the chiral Hermitian matrix ensemble 
\eqref{2}
(this underlying
\eqref{1.1v} in the classification scheme),
\begin{equation}\label{1.3w}
\pi R_{1,\nu}^{\rm NH}(z) =
{d \over dt} \Big ( t \mathcal P[R_{1,\nu}^{\rm H}](2 t) \Big )\Big |_{t = |z|^2}.
\end{equation}
The identity  can be checked since both the complex and Hermitian densities are known explicitly. 
The theory of \cite{CXLR26} gives that \eqref{1.3w} continues to remain valid for all classes of non-Hermitian random matrices in the classification scheme with a rotationally invariant spectrum and which have a (complex) hard edge at the origin. The Hermitian pair can be found within the classification scheme of \cite{AZ97}. The Hermitian hard edge spectral density is known in all cases, whereas apart from the result  \eqref{1.2v} the functional form of the complex densities were not previously known.

\section{Explicit joint eigenvalue PDFs exhibiting a hard edge}\label{S3}
\subsection{Wishart and chiral ensembles}\label{S3.1}
Complex Wishart matrices were specified in the first paragraph of the Introduction, while real Wishart matrices in the special case that the underlying Gaussian matrix is square were introduced in the second paragraph of \S
\ref{S2.1}. For the case that the underlying Gaussian matrix is rectangular, the only difference relative to the construction in the complex case is that the entries are real. Another possibility is to double the size of the Gaussian matrix $X$ from $n \times N$ to $(2n) \times (2N)$ by replacing each independent Gaussian entry by a $2 \times 2$ block representing a quaternion; see e.g.~\cite[Eq.~(1.20)]{Fo10}. The eigenvalues of $W = X^\dagger X$ are then doubly degenerate. The three choices of number fields for the entries real, complex and quaternion are conveniently labelled by the Dyson indices $\beta =1$, $\beta = 2$, $\beta =4$ respectively. They have joint eigenvalue PDF proportional to \cite[Prop.~3.2.2]{Fo10}
\begin{equation}\label{1.1}
\prod_{l=1}^N \lambda_l^{\beta a/2} e^{- \beta \lambda_l/2}
\mathbbm{1}_{\lambda_l > 0} \prod_{1 \le j < k \le N}
| \lambda_k - \lambda_j |^\beta, \quad a =
n - N + 1 - 2/\beta.
\end{equation}
Householder transformations applied to the Gaussian (real or complex) $n \times N$ matrix $X$ can be used to reduce it to an $N \times N$ bidiagonal form
\cite{Si85,DE02}. All the non-zero entries are independently distributed. They are specified by
\begin{equation}\label{1.1a}
(\chi_{\beta n}, \chi_{\beta (n-1)}, \dots,
\chi_{\beta(n-N+1)}), \quad
(\chi_{\beta (N-1)}, \chi_{\beta (N-2)}, \dots,
\chi_{\beta}),
\end{equation}
on the diagonal and lower subleading diagonal respectively.
The notation $\chi_p^2$ refers to the random variable with gamma distribution $\Gamma[p/2,2]$. One notes that this bidiagonal matrix is well defined for continuous $\beta > 0$ and $n \ge N$.
The 2002 result of \cite{DE02} is that after scaling by $1/\sqrt{\beta}$, the squared singular values have the PDF (\ref{1.1}) in such a generalised setting.

The chiral ensembles refer to the matrix structure (\ref{2}), with the rectangular matrix $X$ therein a standard Gaussian matrix as for the construction of the Wishart matrices. It has already been noted that the positive eigenvalues of the chiral matrices, $\{ \mu_i \}_{i=1}^N$, are the positive square roots of the eigenvalues of $X^\dagger X$. Thus the joint eigenvalue PDF of matrices from the chiral ensemble is
proportional to \cite[Prop.~3.1.3]{Fo10}
\begin{equation}\label{1.2}
\prod_{l=1}^N \mu_l^{\beta \alpha} e^{- \beta \mu_l^2/2}
\mathbbm{1}_{\mu_l > 0} \prod_{1 \le j < k \le N}
| \mu_k^2 - \mu_j^2 |^\beta, \quad \alpha =
n - N + 1 - 1/\beta.
\end{equation}
An equivalent interpretation of (\ref{1.2})
for $\beta = 1,2,4$ (continuous $ \beta > 0$) is as the PDF for the singular values of
the Gaussian matrices $X$ (bidiagonal matrices (\ref{1.1a})).

Gaussian ensembles based on the matrix structures (\ref{6b}) admit explicit functional forms \cite[\S 3.2.2]{Fo10}. In the case that the second is of size $4N \times 4N$ as a complex matrix, this and the fourth can be demonstrated to be unitary equivalent. Let us then consider further the second structure, taken to be of size $2N \times 2N$ as a complex matrix. Being a Hermitian matrix based on quaternions, the spectrum is doubly degenerate. In addition, the remaining $N$ eigenvalues come in pairs $\pm \lambda$, implying $m = N/2$ ($N$ even), $m = (N-1)/2$ ($N$ odd) independent (positive) eigenvalues --- for $N$ odd there is also one zero eigenvalue. One has that with independent Gaussian entries, the PDF of the positive eigenvalues is given by (\ref{1.2}) with $N = m$, $\beta = 4$, $\alpha = 1/4$ ($N$ even), $\alpha = 5/4$ ($N$ odd). For $i$ times an $N \times N$ anti-symmetric matrix, which is the first structure in (\ref{6b}), $m$ as defined two sentences above is again the number of independent (positive) eigenvalues, with zero an eigenvalue for $N$ odd. The PDF of the positive eigenvalues in the case of Gaussian entries is given by 
(\ref{1.2}) with $N = m$, $\beta = 2$, $\alpha = 0$ ($N$ even), $\alpha = 1$ ($N$ odd). For the third structure in (\ref{6b}), with $C$ having 
independent complex Gaussian entries up to the required transpose symmetry, the PDF of the positive eigenvalues
is given by (\ref{1.2}) with $\beta = 1$, $\alpha = 1$.

\subsection{Jacobi ensembles}\label{S3.2}
After scaling and renaming of variables, the one-body factors in (\ref{1.1}) are of the form $\lambda^c e^{-\lambda} {\mathbbm 1}_{\lambda > 0}$. This is the Laguerre weight from the theory of orthogonal polynomials. Due to this and keeping in mind too the exponent $\beta$ associated with the product over pairs, (\ref{1.1}) is referred to as the Laguerre $\beta$
ensemble, which in this context is then specified by its eigenvalue probability density. Replacing the Laguerre weight by the Jacobi weight (shifted to be supported on $(0,1)$ for convenience) gives the Jacobi $\beta$ ensemble analogue of (\ref{1.1}),
\begin{equation}\label{1.1J}
\prod_{l=1}^N 
\lambda_l^{\beta \alpha_1/2} (1 - \lambda_l)^{\beta \alpha_2/2}
\mathbbm{1}_{1 > \lambda_l > 0} \prod_{1 \le j < k \le N}
| \lambda_k - \lambda_j |^\beta, \qquad  \alpha_1,  \alpha_2 > - 2/\beta.
\end{equation}
For fixed $\alpha_1, \alpha_2$ there is a hard edge at both 0 and 1, a situation which has already been noted in
\S \ref{S2.3}.

Let $W_1 = X_1^\dagger X_1$, $W_2 = X_2^\dagger X_2$ be two independent Wishart matrices with $X_1$ of size $n_1 \times N$ and $X_2$ of size $n_2 \times N$. Let the entries be real ($\beta = 1$), complex ($\beta = 2$) or quaternion ($\beta = 4$). Form from these the positive definite matrix $W_1(W_1+W_2)^{-1}$, as is relevant to the theory of canonical correlations in multivariate statistics --- see e.g.~\cite{Mu82}. One can compute that the eigenvalue probability density is proportional to
(\ref{1.1J}) with
\begin{equation}\label{1.3}
\alpha_1 = n_1 - N + 1 - 2/\beta, \qquad
\alpha_2 = n_2 - N + 1 - 2/\beta.
\end{equation}
Let $X = [X_1 \: X_2]^T$ be the block matrix of size $(n_1 + n_2)$ rows and $N$ columns. It is readily checked that $W_1^{1/2}(W_1+W_2)^{-1}W_1^{1/2}$ can equivalently be written as the projection matrix $X^\dagger J X (X^\dagger X)^{-1}$ with
$J = {\rm diag}((1)^{n_1},(0)^{n_2})$ (here the notation $(u)^p$ denotes $u$ repeated $p$ times)
\cite{Co05,AEK06}.

Several ensembles related to random unitary matrices also give rise to examples of (\ref{1.1J}). Real orthogonal matrices are the subgroup of unitary matrices having all elements real, denoted ${\rm O}(N)$. They can be further classified according to the value of their determinant --- ${\rm O}^+(N)$  are the subgroup of orthogonal matrices with determinant $+1$, while ${\rm O}^-(N)$ is the set of orthogonal matrices with determinant $-1$. For $N$ odd, at least one of the eigenvalues of matrices from ${\rm O}^+(N)$ must equal $1$, while at least one of the eigenvalues of matrices from ${\rm O}^-(N)$ must equal $-1$. For $N$ even a pair of the eigenvalues of matrices from ${\rm O}^-(N)$ must equal $\pm 1$. All other eigenvalues lie on the unit circle in the complex plane and come in complex conjugate pairs $e^{\pm i \theta}$, $0 \le \theta \le \pi$. 
Now form an ensemble of real orthogonal matrices by choosing them to have Haar measure \cite{Hu97,DF17}.
With $\lambda_j = (1 - \cos \theta_j)/2$ the PDF for the eigenvalues in the upper half plane  is of the form (\ref{1.1J}) with $\beta = 2$, $N = N^*$ and 
\begin{equation}\label{1.3d}
(N^*,a,b) = \left \{ \begin{array}{ll}
(N/2,-1/2,-1/2) & {\rm for \, matrices \, in}
\,\,  O^+(N), \, N \, {\rm even}, \\
((N-1)/2,1/2,-1/2) & {\rm for \, matrices \, in}
\, \, O^+(N), \, N \, {\rm odd}, \\
((N-1)/2,-1/2,1/2) & {\rm for \, matrices \, in}
\,\,  O^-(N), \, N \, {\rm odd}, \\
(N/2-1,1/2,1/2) & {\rm for \, matrices \, in}
\,\,  O^-(N), \, N \, {\rm even};
\end{array} \right.;
\end{equation}
see e.g.~\cite[Prop.~3.7.1]{Fo10}. 

Another prominent subgroup of unitary matrices, applicable with $N \mapsto 2N$ and thus of even size, is that of unitary symplectic matrices ${\rm USp}(2N)$. These matrices also have eigenvalues
in complex conjugate pairs on the unit circle. When chosen with Haar measure, the eigenvalue PDF of the eigenvalues on the upper half unit circle is identical to that for ${\rm O}^-(2N+2)$. 

In relation to Gaussian matrices with real or complex entries, it has been commented below (\ref{1.1}) that Householder transforms can be used to obtain bidiagonal matrices with the special feature that all nonzero entries are independently distributed. The analogous procedure applied to Haar distributed real orthogonal matrices gives a Hessenberg matrix --- all entries below the leading subdiagonal are zero. However, a Hessenberg unitary matrix is fully determined by $N-1$ complex parameters $\{ \alpha_{j-1} \}_{j=1,\dots,N-1}$,
$|\alpha_{j-1}| < 1$, and one parameter $\alpha_{N-1}$, $|\alpha_{N-1}|=1$. For real orthogonal matrices, these parameters are all real and furthermore for $N \mapsto 2N$,
$\alpha_{2N-1}=-1$. When chosen with Haar measure, these parameters have particular independent beta distributed random variables. By an appropriate generalisation of the latter to include a factor of $\beta$, in the variables $\lambda_j = (1 - \cos \theta_j)/2$ it is possible to realise (\ref{1.1J}) for all $\beta > 0$ \cite{KN04}.

Consider now a subblock $A$ of size $n_1 \times n_2$ of a Haar unitary matrix with real ($\beta = 1)$, complex ($\beta = 2)$ or quaternion ($\beta = 4)$ entries. These are equivalently real orthogonal, complex unitary and unitary symplectic matrices respectively, although we know from (\ref{1.3d}) that the label $\beta$ as just specified is not applicable with regards to (\ref{1.1J}).
Provided $N \ge n_1 + n_2$, $n_1 \ge n_2$, the eigenvalue PDF of the positive definite matrix $A^\dagger A$ is well defined, and given by 
(\ref{1.1J}) with \cite{Fo06}
\begin{equation}\label{1.3e}
\alpha_1 = n_1 - n_2 +1 - 2/\beta, \quad
\alpha_2 = N - n_1 - n_2 +
1 - 2/\beta.
\end{equation}

One notes that for $n_1 + n_2 = N$ and $\beta = 2$ the exponent $\alpha_2$ in (\ref{1.3e}) vanishes. There is an analogous effect for the eigenvalues of $A^\dagger A$, where $A$ is
bottom corner $n_1 \times n_2$, $n_1 \ge n_2$ subblock of an $(n_1 + n_2) \times (n_1 + n_2)$ matrix from the circular orthogonal ensemble
(matrices $U^T U$ where $U$ is Haar unitary), and from the circular symplectic ensemble (matrices $U^D U$ where the superscript $D$ denotes the quaternion dual, and $U$ is doubled in size --- then the eigenvalues of $A^\dagger A$ is doubly degenerate). Thus in these settings
the eigenvalue PDF is given by
(\ref{1.1J}), with $\beta =1$ (circular orthogonal ensemble),
$\beta =4$ (circular symplectic unitary ensemble),
$\alpha_1$ again as in (\ref{1.3e}), and $\alpha_2 =0$
\cite[Eq.~(2.9)]{Be97}, \cite[Prop.~3.8.1]{Fo10}.

Unitary analogues of the third and fourth class of matrices in (\ref{6b}) are matrices $S S^T$, and
$S S^D$, where $S$ is from ${\rm Sp}(2N)$, and ${\rm O}^+(2N)$. These are unitary equivalent to symmetric and anti-symmetric real orthogonal matrices respectively. In the second of these the eigenvalues are doubly degenerate, with $\lambda = -1$ being such an eigenvalue.
For the former ensemble with Haar measure, the 
PDF of the eigenvalues on the upper half unit circle is, in the
variables  $\lambda_j = (1 - \cos \theta_j)/2$, given by
(\ref{1.1J}) with $\alpha_1 = \alpha_2 = 0$ and $\beta =1$. For the ensemble of anti-symmetric real orthogonal matrices, set $m = N/2$ ($N$ even), $m = (N-1)/2$ ($N$ odd). Then with the same change of variables, the 
PDF of the eigenvalues on the upper half unit circle is given by 
(\ref{1.1J}) with $N = m$, $\beta =4$ and $\alpha_1 = \alpha_2 = 0$ for $N$ even, and $\alpha_1 = 1/2$,
$\alpha_2 = 0$ for $N$ odd; see 
\cite{Du04}, \cite[\S 3.7.3]{Fo10}, \cite{B+11}.

\subsection{Brownian motion and parameter dependent ensembles}\label{S3.3}
The chiral matrices of \S\ref{S2.2}, and the Wishart matrices of \S\ref{S3.1}, are formed from rectangular Gaussian matrices $X$ with zero mean. Suppose instead 
a nonzero, parameter dependent, mean is imposed by choosing the matrix distribution of $X$ to be proportional to
\begin{equation}\label{1.6a}
\exp \Big ( - \beta {\rm Tr} \Big \{ \, (X - e^{-\tau}
X^{(0)})^\dagger (X - e^{-\tau}
X^{(0)})  \Big \} / 2 | 1 - e^{-2 \tau} | \Big ),
\end{equation}
where $\beta =1,2$ or 4 distinguishes the number field.
Thus as $\tau \to 0^+$, $X$ is equal to $X^{(0)}$, while as $\tau \to \infty$, $X$ is again a zero mean Gaussian matrix. In the chiral matrix context of the nonzero positive eigenvalues of \eqref{2}, the $\tau$-dependent joint eigenvalue PDF can be shown to satisfy a Fokker-Planck equation (a single element undergoes Brownian motion according to an Ornstein-Uhlenbeck process), closely related to that characterising Dyson Brownian motion for Gaussian Hermitian random matrices \cite{FNH99}. The Fokker-Planck operator is non-Hermitian, but upon a simple similarity transformation it can, for general $\beta > 0$, be mapped to a many body Schr\"odinger operator of Calogero-Sutherland type (inverse square pair potential), which in turn permits a Jack polynomial based multivariate hypergeometric function solution (depending on both the singular values of $X$ and $X^{(0)}$) \cite{BF97a}. The increase in complexity relative to the joint eigenvalue PDF \eqref{1.2} is evident from the fact that in a change of variables from the matrix to its singular values and the singular vectors, the latter do not separate. For $\beta =1,2$ and 4 this means that
\eqref{1.2} is to be multiplied by a certain group integral (see \cite[Eq.~(13.147)]{Fo10} and (\ref{17.ty2}) below for $\beta = 2$) up to scaling of the variables and multiplication by a Gaussian in the singular values of $X^{(0)}$. In the mapping to the many body Schr\"odinger operator, the case $\beta =2$ is special as then the coefficient of the inverse square pair potential vanishes, and one is left with determining the Green function solution for free fermions, which has a well known determinantal structure. Thus, with 
$d^{\rm H} U, \, d^{\rm H} V$ the Haar measure on $U(N), U(n)$,
and $\{x_j^2 \}$, $\{y_j^2\}$ the eigenvalues of $X^\dagger X$, $Y^\dagger Y$, the required group integral can be evaluated \cite{BK58,GW96}, \cite[Prop.~11.6.2]{Fo10}
\begin{equation}\label{17.ty2}
\int  d^{\rm H} U \int d^{\rm H} V \,
e^{{\rm Tr}({ U} { X}^\dagger { V}^\dagger { Y} +
{ Y}^\dagger { VXU}^\dagger)} 
\propto 
{\det [  {}_0 F_1(n-N+1; (x_jy_k)^2) ]_{j,k=1,\dots,N} \over
\prod_{1 \le j < k \le N} (y_k^2 - y_j^2) (x_k^2 - x_j^2) }
\end{equation}
(here the matrix $Y$ is to be identified with $X^{(0)}$ up to proportionality). Consequently, with $X = X^{\rm G} + X^{(0)}$, where $X^{\rm G}$ is an $n \times N$ ($n \ge N$) standard complex Gaussian matrix, and $X^{(0)}$ is a fixed matrix, upon combining (\ref{17.ty2}) and \eqref{1.1} we have that the eigenvalue PDF of $X^\dagger X$ is, with $a=n-N$, proportional to \cite[Prop.~5]{DF08}
\begin{equation}\label{1.1aa}
\prod_{l=1}^N \lambda_l^{a} e^{- \lambda_l}
\mathbbm{1}_{\lambda_l > 0} \prod_{1 \le j < k \le N}
( \lambda_k - \lambda_j )
\det [  {}_0 F_1(a+1; \lambda_j c_k) ]_{j,k=1,\dots,N}.
\end{equation}
Here $\{ \lambda_j \}$ denotes the eigenvalues of $X^\dagger X$, $\{ c_k \}$ denotes the eigenvalues of
$(X^{(0)})^\dagger X^{(0)}$, and the (not shown) proportionality is dependent on $\{ c_k \}$. We refer to
\cite{As23} for a recent treatment.

The above generalisation of the complex Wishart matrix introduced in \S \ref{S3.1}, with the mean now being nonzero, is termed noncentral. Another class of generalisation is to replace $X^{\rm G}$ by $X = X^{\rm G}
\Sigma^{1/2}$, where $\Sigma$ is a positive definite matrix which specifies the covariance. After changing variables to $W = X^\dagger X$ in the corresponding  matrix distribution (i.e.~the generalisation of \eqref{1.6a}), then changing variables to the eigenvalues and eigenvectors of $W$, the matrix group integral named after Harish-Chandra/ Itzykson--Zuber (often abbreviated HCIZ) is encountered,
\begin{equation}\label{17.HHa}
\int_{{\rm U}(N)} e^{{\rm Tr}({ UAU}^\dagger { B})} \, d^{\rm H} U =
\prod_{j=1}^{N-1}j! \, {\det [ e^{a_j b_k} ]_{j,k=1,\dots,N} \over
\prod_{1 \le j < k \le N} (a_k - a_j) (b_k - b_j) };
\end{equation}
see e.g.~\cite[Prop.~11.6.1]{Fo10}. Here $\{a_j\}$,
$\{b_j\}$ are the eigenvalues of $A$, $B$. This leads to the result that the eigenvalue PDF of $W$ in this setting is proportional to \cite{GS00}
\begin{equation}\label{17.HHb}
\prod_{l=1}^N \lambda_l^{a} e^{- \lambda_l}
\mathbbm{1}_{\lambda_l > 0} \prod_{1 \le j < k \le N}
( \lambda_k - \lambda_j )
\det [e^{-\lambda_j c_k}]_{j,k=1,\dots,N},
\end{equation}
where $\{1/(1+c_k)\}$ are the eigenvalues of the positive definite matrix $\Sigma$, and the proportionality is dependent on $\{c_k\}$. 

There is a special feature of the case $\Sigma = {\rm diag}(b,1,\dots,1)$, $b>0$, which corresponds to a multiplicative rank 1 perturbation of $X^{\rm G}$. Thus, since the nonzero eigenvalues of
$\Sigma^{1/2} (X^{\rm G})^\dagger X^{\rm G} \Sigma^{1/2}$ are the same as for $X^{\rm G} \Sigma (X^{\rm G})^\dagger$, one has that for this case the eigenvalue probability density is the same as that for the additive rank 1 perturbation
$X_1^{\rm G} (X_1^{\rm G})^\dagger + b \mathbf x \mathbf x^\dagger$, where $X_1^{\rm G}$ refers to $X^{\rm G}$ with the first column deleted, and $\mathbf x$ denotes the first column of $X^{\rm G}$; see e.g.~\cite[\S 3.1]{Fo23}.
This setting also leads to a rank 1 generalisation of the bidiagonal matrix \eqref{1.1a}, which we recall has (after scaling) the functional form \eqref{1.1} for the PDF of its squared singular values. Thus one multiplies the first entry in the first of the arrays in \eqref{1.1a} by $\sqrt{b}$, and leaves all the other entries unchanged \cite{Fo13}. The generalisation of \eqref{1.1} is then the extra multiplicative factor
\begin{equation}\label{17.HHc}
\int_{-\infty + i c}^{\infty + i c} e^{it}
\prod_{j=1}^N \Big ( i t - {b - 1 \over 2 b} \beta \lambda_j \Big )^{-\beta/2} \, dt,
\end{equation}
where $c \ge 0$ is chosen so that the branch of the square root is to the left. In the case $\beta =2$ this can be checked to be consistent with the appropriate specialisation of \eqref{17.HHb}.

The notion of Brownian motion on the unitary group was introduced into random matrix theory in the 1962 work \cite{Dy62b}, leading to a parameter dependent theory of the eigenvalues. This construction, based on the infinitesimal equation $S(\tau + \delta \tau) = S(\tau) e^{i \sqrt{\delta \tau} M(\tau)}$ with $M(\tau)$ drawn from  the GOE $(\beta = 1)$, GUE $(\beta = 2)$ or
GSE $(\beta = 4)$ independently for each $\tau$, applies to all three circular ensembles.
It was shown in \cite{FP95,KP10} that now decomposing such parameter dependent unitary matrices according 
to \eqref{4}, a Fokker-Planck operator can be identified specifying the evolution of the joint eigenvalue PDF  for $T^\dagger T$.
With the eigenvalues parameterised as 
$\{ \sin^2 \phi_j\}_{j=1}^N$, $0< \phi_j < \pi/2$, and upon specialising to the case $n=N$ for simplicity, this leads to the functional form
\begin{equation}\label{1.1bb}
\prod_{l=1}^N \sin (2 \phi_l)
 \prod_{1 \le j < k \le N}
( \sin^2\phi_k - \sin^2 \phi_j )
\det \Big [  \sum_{n=0}^\infty P_n(\cos(2 \phi_j))
P_n(\cos(2\phi_k^{(0)})) e^{-(2n+1)^2 \tau}
\Big ]_{j,k=1,\dots,N},
\end{equation}
where $P_n(x)$ denotes the Legendre polynomial, up to a $\{ \phi_k^{(0)} \}$ dependent normalisation. The variables $\{ \sin^2 \phi_j^{(0)}\}_{j=1}^N$ are the eigenvalues at $\tau = 0$ (these are prescribed), with the (not shown) proportionality dependent on these variables as well as the parameter $\tau$.
Brownian motion models can similarly be formulated for classical real orthogonal, and unitary symplectic groups
(see e.g.~\cite[\S 7]{CMO18}). For example, in the case of ${\rm O}^+(2N)$, replacing (\ref{1.1bb}) is
\begin{equation}\label{1.1cc}
 \prod_{1 \le j < k \le N}
( \cos \phi_k - \cos \phi_j )
\det \Big [  \sum_{n=-\infty}^\infty \cos n \phi_j
\cos n\phi_k^{(0)} e^{-n^2 \tau}
\Big ]_{j,k=1,\dots,N}.
\end{equation}

Considered in the 2019 work \cite{AKMV19} was the parametric sum
\begin{equation}\label{1.1dd}
a i A + i (1 - a^2)^{1/2} \begin{bmatrix} 
0_N & W \\
- W^T & 0_{N} \end{bmatrix}, \quad 0 \le a  \le 1.
\end{equation}
The matrix $A$ is a $2N \times 2N$ standard real Gaussian anti-symmetric matrix, while $W$ is an $N \times N$ standard real Gaussian matrix. The eigenvalues of \eqref{1.1dd} come in $\pm$ pairs. In the case $N$ even (for definiteness) it was shown that the PDF for the positive eigenvalues is proportional to
\begin{equation}\label{1.1ee}
\prod_{l=1}^N e^{-2 \lambda_l^2} \prod_{1 \le j < k \le N} ( \lambda_k^2 - \lambda_j^2)
{\rm Pf} \Big [ {\rm erf}(\gamma(\lambda_k - \lambda_j))
{\rm erf}(\gamma(\lambda_k + \lambda_j)) \Big ]_{j,k=1}^N,
\end{equation}
with $\gamma = \sqrt{(1-a^2)/a^2}$. Crucial to its derivation is the version of the HCIZ integral relating to anti-symmetric matrices and the orthogonal group
\begin{equation}\label{1.1ff}
\int_{{\rm O}(2N)} e^{{\rm Tr} (  X R  Y R^T/2 )} \, d^{\rm H} R =
c_N \, {\det [ 2 \cosh (x_j y_k) ]_{j,k=1,\dots,N} \over
\prod_{1 \le j < k \le N} (x_k^2 - x_j^2) (y_k^2 - y_j^2) }, \quad  c_N = \prod_{j=1}^N {\Gamma(2j-1) \over 2},
\end{equation}
for $X,Y$ real anti-symmetric matrices with eigenvalues
$\{\pm i x_j \}_{j=1}^N$, $\{\pm i y_j \}_{j=1}^N$;
see e.g.~\cite{Mc21}.

As a further parametric sum consider $M = \sqrt{(1 + \tau)/2} H_1 + i \sqrt{(1-\tau)/2} H_2$, 
$|\tau| \le 1$, where $H_1, H_2$ are $N \times N$ matrices from the GUE. These non-Hermitian matrices form the elliptic Ginibre ensemble --- see e.g.~\cite{BF25}.
In the 2018 work \cite{KK18},
by providing an evaluation of the unitary matrix integral
\begin{equation}\label{1.1gg}
\int_{{\rm U}(N)}
\exp \Big (  {\rm Tr} ((XU)^2 + (U^\dagger X)^2) \Big )
\, d^{\rm H} U
\end{equation}
(see also \cite{LW26,AAW24}), it  was shown 
that the probability density for the squared singular values of $M$
is proportional to
\begin{equation}\label{1.1hh}
\prod_{l=1}^N e^{-(\eta_+ + \eta_-) \lambda_l} \prod_{1 \le j < k \le N}
(\lambda_k - \lambda_j)
{\rm Pf} \Big [ \mathcal E(\eta_- \lambda_j, \eta_- \lambda_k) \Big ]_{j,k=1}^N.
\end{equation}
Here, 
$N$ is assumed even for convenience,
and in the notation from \cite{LW26},
$\eta_+ = 1/(1 - \tau^2)$,
$\eta_- = \tau/(1 - \tau^2)$ and
\begin{equation}\label{1.1jj}
\mathcal E(u,v) = \int_{\mathbb R^2} dx d y \,
{x - y \over x + y} e^{- {1 \over 2} (x^2 + y^2)}
\Big ( I_0(2 x \sqrt{u}) I_0(2 y \sqrt{v}) -
I_0(2 y \sqrt{u}) I_0(2 x \sqrt{v})  \Big ).
\end{equation}
Included in this model is the case of the (squared) singular values for GUE matrices. This is obtained by setting $\tau = 1$ in the original definition of the matrix $M$ --- note that the limit $\tau \to 1^-$ is singular in \eqref{1.1hh}. Denote the distribution of the GUE singular values by $|{\rm GUE}_N|$ and the result of applying the change of variables $\lambda_l \mapsto \lambda_l^2$ to the LUE eigenvalue PDF specified by \eqref{1.1} with $\beta=2$ by $\widetilde{\rm LUE}_{N,a}$. For $N$ even one has the identity in distribution
\cite{Fo06x,EL15}
\begin{equation}\label{1.1kk}
| {\rm GUE}_N|
\mathop{=}\limits^{\rm d} 
\widetilde{\rm LUE}_{N/2,-1/2}
\cup 
\widetilde{\rm LUE}_{N/2,1/2},
\end{equation}
where $N$ has been chosen as even for convenience.

\subsection{Positive definite product ensembles}\label{S3.4}
The result of the second paragraph of \S \ref{S3.2} for the eigenvalue PDF of $W_1 ( W_1 + W_2)^{-1}$ gives, upon the change of variables $\lambda_j \mapsto 1/(1 + \lambda_j)$, the eigenvalue 
PDF of the product $W_1^{-1} W_2$. Thus one sees that (\ref{1.1J}) maps to
\begin{equation}\label{1.4a}
\prod_{l=1}^N {\lambda_l^{\beta \alpha_2/2} \over
(1 + \lambda_l)^{\beta(N-1 + 2/\beta + (\alpha_1 + \alpha_2)/2) }} \mathbbm{1}_{\lambda_l > 0}
\prod_{1 \le j < k \le N}
| \lambda_k - \lambda_j |^\beta,
\end{equation}
in which we then substitute (\ref{1.3}). This is sometimes referred to as the Jacobi prime ensemble; see e.g.~\cite{FI18}. One remarks that the hard edge at $\lambda =1$ in (\ref{1.1J}) maps  to a hard edge at $\lambda = 0$ in (\ref{1.4a}), while the hard edge at $\lambda = 0$ maps to infinity in (\ref{1.4a}).

Beyond this result for $W_1^{-1} W_2$ as applicable for the three number fields corresponding to real, complex, or quaternion entries, further progress along these lines requires restriction to complex entries $\beta = 2$). 
Then the eigenvalue PDF for both the Laguerre and Jacobi ensembles can be written in the so-called P\'olya ensemble form \cite{FKK21} (see too \cite{Ku16,KK19})
\begin{equation}\label{1.4b}
\prod_{1 \le j < k \le N} (x_k - x_j) 
\det \Big [ \Big ( - x_k {\partial \over \partial x_k}
\Big )^{j-1} w(x_k) \Big ]_{j,k=1}^N.
\end{equation}
To see this one makes use of the identity
\begin{equation}\label{1.4c}
\det \Big [ \Big ( - x_k {\partial \over \partial x_k}
\Big )^{j-1} w(x_k) \Big ]_{j,k=1}^N =
\det \Big [ \Big ( {\partial \over \partial x_k}
\Big )^{j-1} ( - x_k)^{j-1} w(x_k) \Big ]_{j,k=1}^N.
\end{equation}
With $w(x)$ given by the Laguerre and Jacobi weights
\begin{equation}\label{1.4d}
w^{\rm L}(x) = x^a e^{-x} {\mathbbm 1}_{x > 0}, \qquad
w^{\rm J}(x) = x^{\alpha_1}  (1 - x)^{N+\alpha_2-1} {\mathbbm 1}_{1 > x > 0},
\end{equation}
use of the Rodrigues formula for the Laguerre and Jacobi polynomials, and the Vandermonde determinant formula, gives (\ref{1.1}) and (\ref{1.1J}) respectively for $\beta = 2$. The utility of the structure (\ref{1.4b}) is that if $Y_1$ ($Y_2$) have this structure with weight $w_1$ ($w_2$), 
then so too does the product matrix $Y_1 Y_2$ with weight
\begin{equation}\label{1.4e}
w(x) = \int_0^\infty dx_1 \int_0^\infty dx_2 \,
\delta(x - x_1 x_2) w_1(x_1) w_2(x_2),
\end{equation}
as well as the product matrix $X_1X_2^{-1}$ with weight
\begin{equation}\label{1.4f}
w(x) = \int_0^\infty dx_1 \int_0^\infty dx_2 \, x_2^{N+1}
\delta(x - x_1/ x_2) w_1(x_1) w_2(x_2).
\end{equation}
Note that the latter with two distinct Laguerre weights can be evaluated to give the $\beta = 2$ Jacobi prime weight in (\ref{1.4a}).
Generally the integrals (\ref{1.4e}) and
(\ref{1.4f}), and higher order iterates, can be conveniently simplified using the Mellin transform. This in turn leads the evaluation of the eigenvalue PDF for products of Wishart matrices and their inverses, and for the products of the squared singular values of truncations of Haar unitary matrices and their inverses, in terms of particular Meijer G-functions; see the review \cite{AI15}, as well as the original works
\cite{AIK13,KS14,KKS16}. One remarks that an equivalent viewpoint of these eigenvalue PDFs is that of the scaled singular values of the product (of compatible sizes) of the underlying non-Hermitian matrices.

In the case of a product of $m$ Laguerre ensembles with exponents $a_1,\dots, a_m$, the Mellin transform strategy and the definition of the Meijer G-function gives
\begin{equation}\label{1.4g}
w(x) = \prod_{j=1}^m {1 \over \Gamma(a_j+1)}
\MeijerG{m}{0}{0}{m}{\underline{\hspace{0.5cm}}}{a_1, \dots, a_m}{x} {\mathbbm 1}_{x > 0};
\end{equation}
see e.g.~\cite[Eq.~(3.3)]{FL22}.
With $a_j = c + (j-1)/m$ $(j=1,\dots,m)$, this Meijer G-function simplifies, giving a functional form proportional to
\begin{equation}\label{1.4h}
x^c e^{-m x^{1/m}} {\mathbbm 1}_{x > 0}.
\end{equation}
Substituting in (\ref{1.4b}) and simplifying, then changing variables
$m x_l^{1/m} =
 \lambda_l$ gives, upon redefinition of $c$, an
eigenvalue PDF proportional to
 \begin{equation}\label{1.4j}
 \prod_{l=1}^N \lambda_l^c e^{- \lambda_l }
 \mathbbm{1}_{\lambda_l > 0}
\prod_{1 \le j < k \le N} (\lambda_k - \lambda_j)
 (\lambda_k^{m} - \lambda_j^{m}).
 \end{equation}
 This generalisation of the $\beta = 2$ case of \eqref{1.1}
specifies the Laguerre Muttalib--Borodin ensemble \cite{Mu95,Bo98}. 
In the case $m=2$, $c = 2 (N - n)$ it is the
eigenvalue PDF of the positive eigenvalues for the Hermitian anti-symmetrix random matrix (eigenvalues occur in pairs 
$\pm \lambda$)
 \begin{equation}\label{1.4ja}
 X \bigg ( \mathbb I_N \otimes \begin{bmatrix} 0 & i \\
 -i & 0 \end{bmatrix} \bigg ) X^T,
 \end{equation}
 where $X$ is a $2n \times 2N$ standard real Gaussian matrix \cite{LSZ06,FILZ19}.
More generally, \eqref{1.4j} permits a random matrix realisation for all $m > 0$ \cite{Ch18,FW17}. In the limit $m \to 0$ \eqref{1.4j} is proportional to
\begin{equation}\label{1.4k}
 \prod_{l=1}^N \lambda_l^c e^{- \lambda_l }
 \mathbbm{1}_{\lambda_l > 0}
\prod_{1 \le j < k \le N} (\lambda_k - \lambda_j)
 (\log \lambda_k - \log \lambda_j).
 \end{equation}
 With the unit trace constraint $\sum_{l=1}^N \lambda_l =
 1$, this joint eigenvalue density has shown itself in recent studies in quantum information \cite{Mi25,SW26}.

 Let $Y$ be an $N \times N$ random matrix such that $Y^\dagger Y$ has an eigenvalue PDF proportional to 
 \begin{equation}\label{1.5a}
 \prod_{1 \le j < k \le N} (x_k - x_j) 
 \det [ f_{k-1}(x_j) ]_{j,k=1}^N.
  \end{equation}
  Let $(X^{\rm G})^\dagger X^{\rm G}$ be a complex Wishart matrix. Closely related to the theory of the paragraph including (\ref{1.4b}) is the fact that the PDF of the eigenvalues of $(X^{\rm G} Y)^\dagger 
  X^{\rm G}  Y$, or equivalently of the eigenvalues of the square singular values of $X^{\rm G} Y$, is also of the form 
  (\ref{1.5a}) with $f_l(y)$ replaced by \cite{KS14}
  \begin{equation}\label{1.5b}
  \int_0^\infty x^a e^{-x} f_l\Big ( {y \over x} \Big ) \, {dx \over x}.
 \end{equation}
 An application given in \cite{FL16} is to obtain the explicit $\{ f_{k-1}(c) \} $ in (\ref{1.5a}) corresponding to the eigenvalue PDF of the product
  \begin{equation}\label{1.5c}
  (X_m^{\rm G})^\dagger X_m^{\rm G} \cdots
  (X_1^{\rm G})^\dagger X_1^{\rm G}
  (X^{\rm G}  + X^{(0)})^\dagger
   (X^{\rm G}  + X^{(0)}).
\end{equation}
This then is a generalisation of 
\eqref{1.1aa}, and of \eqref{1.4b} with substitution
\eqref{1.4g}.

A PDF structure more general than \eqref{1.5a} is
\begin{equation}\label{1.5at}
\det [ g_{k-1}(x_j) ]_{j,k=1}^N
 \det [ f_{k-1}(x_j) ]_{j,k=1}^N,
  \end{equation}
  said to specify a biorthogonal ensemble \cite{Bo98}
  (see too the recent work \cite{CZ26}). With $A,B$ independent rectangular standard complex Gaussian matrices, the singular values of the parameter dependent (parameter $\mu$) matrix product
  $(i A + \mu B)(i A^\dagger + \mu B^\dagger)$ was shown to have PDF of the form \eqref{1.5at} in \cite{AS16}.

A product ensemble with PDF for its squared singular values distinct from the structure
\eqref{1.5at} is $(\mathbb I + U) X^{\rm G}$, where $U$ is Haar unitary, or more generally from the circular unitary ensemble \cite{OSZ10,FK16}. Up to proportionality, this PDF reads
\begin{equation}\label{1.4ks}
 \prod_{l=1}^N \lambda_l^{\alpha - 1/2} 
 e^{- \lambda_l }
 \mathbbm{1}_{\lambda_l > 0}
\prod_{1 \le j < k \le N} {(\lambda_k - \lambda_j)^2 \over \lambda_k + \lambda_j}.
 \end{equation}
 It was first isolated in relation to specifying a statistical distribution on density matrices for finite dimensional random bipartite quantum systems
 \cite{Ha98,SZ04}. By virtue of the Schur Pfaffian identity \cite{IOTZ95}
 \begin{equation}\label{1.4kt}
 \prod_{1 \le j < k \le N} {\lambda_k - \lambda_j \over \lambda_k + \lambda_j}
 = {\rm Pf} \bigg [{\lambda_k - \lambda_j \over \lambda_k + \lambda_j} \bigg ]_{j,k=1}^N,
 \end{equation}
 where it has been assumed that $N$ is even for convenience, one observes that 
 \eqref{1.4ks} can be written in a form analogous to \eqref{1.1ee} and
 \eqref{1.1hh}, which is that specifying a Pfaffian point process. In fact \eqref{1.4kt} is also an ingredient in the derivation of \eqref{1.1hh} \cite{KK18}.

 \section{Optimal hard edge scaled large $N$ expansions for the JUE}

 \subsection{Introductory remarks}
Consider first complex Wishart matrices, which we know to have joint eigenvalue PDF \eqref{1.1} with $\beta = 2$, this also being referred to as the Laguerre unitary ensemble (LUE). It was remarked in the opening paragraph of the Introduction that the hard edge scaling is $\lambda \mapsto \lambda/4 N$, this then resulting in a well defined limiting statistical state (with spacings between eigenvalues of order unity) in the neighbourhood of the origin.
This statistical state is a determinantal point process, and so determined by a single correlation kernel (the so-called Bessel kernel \cite{Fo93}),
 \begin{equation}\label{3.1}
K^{\rm h}_{\infty,0}(x,y) = 
{ J_{a}(x^{1/2}) 
 y^{1/2} J_{a}'(y^{1/2}) - 
 J_{a}(y^{1/2}) 
 x^{1/2} J_{a}'(x^{1/2}) \over 2 (x - y)} \mathbbm 1_{x,y>0}.
 \end{equation}
Now set
\begin{equation}\label{3.1a}
N_{\rm h}' = N + a/2
\end{equation}
and consider the correlation kernel for the LUE.
Recently established in \cite{FRS26} is a large $N$ expansion in powers of $(N_{\rm h}')^{-2}$,
\begin{equation}\label{3.1b}
{1 \over 4N_{\rm h}'}
K_N^{\rm LUE}(x/4 N_{\rm h}',
y/4 N_{\rm h}') =
K^{\rm h}_{\infty,0}(x,y) +
{1 \over (N_{\rm h}')^2}
K^{\rm h,L}_{\infty,1}(x,y) +
{1 \over (N_{\rm h}')^4}
K^{\rm h,L}_{\infty,2}(x,y) + \cdots,
\end{equation}
for certain functions $
K^{\rm h,L}_{\infty,j}(x,y)$, $(i=1,2,\dots)$,
which moreover represents the optimal rate of convergence. At the order of the first correction, the variable \eqref{3.1a} was already identified in several earlier works
\cite{EGP16,Bo16,PS16,HHN16,FT19}. Preceding \eqref{3.1b} was the discovery of analogous expansions at the soft edge for Gaussian and Laguerre $\beta$ ensembles for the classical values $\beta = 1,2$ and 4, although in the first instance the quantity being expanded was (the generating function for) the conditioned soft edge gap probability \cite{Bo24,Bo25a}, and later the density
\cite{Bo25,FRS26}. In this setting the expansions proceed in inverse powers of $N_{\rm s}':= N + (\beta - 2)/(2 \beta)$, which is thus a $\beta$ modification of $N$. Actually at $\beta =2$ the method of study proceeds via soft edge analogues of the expansion \eqref{3.1b}; see also \cite{YZ26,LMZ26}.

In this Section we carry out a study relating to optimal large $N$ expansions at the hard edge for the Jacobi ensemble \eqref{1.1J}, although we find it preferable to take the exponents in the weight functions therein to be independent of $\beta$ by considering
\begin{equation}\label{WJ}
w^{\rm J}(\lambda) := \lambda^a (1 - \lambda)^b \mathbbm 1_{0 < \lambda < 1}.
\end{equation}
It has already been remarked that the conventional hard edge scaling for the Laguerre ensemble
\eqref{1.1} is  $\lambda \mapsto \lambda/4 N$. To anticipate the analogue for the Jacobi ensemble, one first notes that the leading order support of the Laguerre ensemble is $(0,4N)$; see e.g.~\cite[Prop.~3.2.3]{Fo10}. Hence the spacing between bulk eigenvalues is of order unity, and those at the hard edge is of order $1/N$.
For the Jacobi ensemble the spacing between bulk eigenvalues is of order $1/N$, leading us to anticipate that the edge eigenvalues have spacing of order $1/N^2$.
Indeed the Jacobi hard edge scaling
$\lambda \mapsto \lambda/ 4 N^2$ to give agreement with the hard edge scaling limit for the Laguerre ensemble is known from other considerations \cite[\S 7.2.5]{Fo10}.
Also, previous work \cite[Appendix A, arXiv version]{FL22},
\cite{Wi24,MMM19} indicates that the $\beta$ generalised Jacobi analogue of the optimal expansion variable (\ref{3.1a}) using the Jacobi weight \eqref{WJ} is
\begin{equation}\label{3.1J}
N_{\rm hJ}' = N - 1/2 + (a+b+1)/\beta
\end{equation}
(the $\beta$ generalisation of (\ref{3.1a}) in the Laguerre ensemble with $\beta$ independent Laguerre exponent $a$
is $N_{\rm h}'=N + a/\beta$ \cite{FT19}).

 \subsection{The JUE correlation kernel}
 The correlation kernel associated with the determinantal structure implied by \eqref{1.1J} with $\beta = 2$ --- this is said to specify the Jacobi unitary ensemble (JUE) --- has the Christoffel--Darboux form
 \begin{equation}\label{3.2J}
 K_N^{\rm JUE} (x,y) =
 \Big ( x^a (1 - x)^b y^a (1 - y)^b \Big )^{1/2} {1 \over h_{N-1}}
 {p_N(x) p_{N-1}(y) - p_N(y) p_{N-1}(x) \over x - y},
 \end{equation}
 where 
 \begin{equation}\label{3.3J}
p_n(x) = (-1)^n {(a+1)_n \over (a + b + n + 1)_n}
{}_2 F_1(-n,a+b+n+1,a+1;x)
  \end{equation}
and
 \begin{equation}\label{3.4J}
h_n = {n! \Gamma(a+n+1) \Gamma(b+n+1) \Gamma(a+b+n+1) \over (a + b + 2n + 1)
\Gamma^2(a+b+2n+1) }
\end{equation}
(see e.g.~\cite[Eq.~(2.26) and (2.27), the latter after correction]{Fo21}, with
\eqref{3.3J} being a hypergeometric form of a (shifted) Jacobi polynomial \cite{CS95}). From the series definition of the hypergeometric polynomial, we see from this that a hard edge scaling of \eqref{3.2J} gives rise to a large $N$ expansion in powers of $1/N$, as a formal series at least (it does not provide an error bound upon truncation of the series). In fact use of the shifted variable $N_{\rm hJ}'$ \eqref{3.1J} with $\beta =2$ for the hard edge scaling in place of $N$ shows that the expansion proceeds in powers of $1/(N_{\rm hJ}')^2$.

\begin{prop}\label{P4.1}
We have that
 \begin{equation}\label{3.5J}
  {1 \over 4 (N_{\rm hJ}' )^2} K_N^{\rm JUE} \bigg ( {x \over 4 (N_{\rm hJ}' )^2},
  {y \over 4 (N_{\rm hJ}' )^2} \bigg ),
  \end{equation}
defined as a continuous function of $N_{\rm hJ}'$ by
(\ref{3.2J})--(\ref{3.4J}), is even in this variable for non-negative integer $a$. This property remains true of the large $N_{\rm hJ}'$ expansion for continuous $a$. 
\end{prop}

\begin{proof}
We have
{\small \begin{multline} 
{}_2 F_1(-n,a+b+n+1,a+1;x) \Big |_{\substack{n \mapsto N \\
N_{\rm hJ}' \mapsto - N_{\rm hJ}'}} =
{}_2 F_1(N_{\rm hJ}'+(a+b)/2,(a+b)/2-N_{\rm hJ}'+1,a+1;x) \\
= {}_2 F_1((a+b)/2-N_{\rm hJ}'+1,N_{\rm hJ}'+(a+b)/2,a+1;x) =
{}_2 F_1(-n,a+b+n+1,a+1;x) \Big |_{n \mapsto N-1},
\end{multline}}
where for the second equality the general symmetry of the
hypergeometric function ${}_2 F_1(a,b,c;x) =
{}_2 F_1(b,a,c;x)$ has been used.
This shows that  the terms dependent on $x,y$ 
in \eqref{3.2J} are odd in $N_{\rm hJ}'$. It remains to consider the terms independent of $x,y$,
 \begin{multline}\label{3.6J}
-{1 \over h_{N-1}}
{(a+1)_N \over (a + b + N + 1)_N} 
{(a+1)_{N-1} \over (a + b + N)_{N-1}} \\ = 
-{1 \over 2 N_{\rm hJ}' \Gamma^2(a+1)} {\Gamma( N_{\rm hJ}' + 1 - (b-a)/2) \over \Gamma( N_{\rm hJ}' - (a+b)/2)  }
{\Gamma( N_{\rm hJ}' + 1 + (a+b)/2) \over \Gamma( N_{\rm hJ}' + (b-a)/2)  }.
 \end{multline}
 For $a$ a non-negative integer, use of the recurrence for the gamma function exhibits that each of the final two factors (both ratios) are even polynomials in $N_{\rm hJ}'$. Hence \eqref{3.6J} is odd in $N_{\rm hJ}'$, which combined with the established fact that  
the terms dependent on $x,y$ 
in \eqref{3.3J} are odd in $N_{\rm hJ}'$ gives the first stated result. For the second stated result, we use the fact that the coefficients in the asymptotic expansion of the ratio of gamma functions are polynomials in $a,b$
\cite{TE51}, and thus agree with those for $a$ integer for $a$ large enough. In particular, the large $N_{\rm hJ}'$ expansions must agree.
Thus, after dividing through by $(N_{\rm hJ}')^{2a}$ as results from the substitutions required by
\eqref{3.5J} in the factors $(x^ay^a)^{1/2}$ of \eqref{3.2J}, the expansion must 
be even in this variable, as is readily illustrated to low order to be
\begin{equation}\label{3.7}
 (N_{\rm hJ}')^2 \bigg (  1 -
{1 \over 12} {(a+1)(2a + a^2 + 3 b^2) \over  (N_{\rm hJ}')^2}
   + \cdots   \bigg ).
   \end{equation}
\end{proof}

Next we would like to investigate the functions on the right hand side of the analogue of \eqref{3.1b} for the JUE,
\begin{equation}\label{3.7b}
 {1 \over 4 (N_{\rm hJ}' )^2} K_N^{\rm JUE} \bigg ( {x \over 4 (N_{\rm hJ}' )^2},
  {y \over 4 (N_{\rm hJ}' )^2} \bigg ) =
K^{\rm h}_{\infty,0}(x,y) +
{1 \over (N_{\rm hJ}')^2}
K^{\rm h,J}_{\infty,1}(x,y) +
{1 \over (N_{\rm hJ}')^4}
K^{\rm h,J}_{\infty,2}(x,y) + \cdots,
\end{equation}
the existence of which is implied by Proposition \ref{P4.1} and the existence of an expansion in $1/N$ (for the latter recall the text below \eqref{3.4J}).
That at leading order the right hand sides of \eqref{3.7b} and \eqref{3.1b} are the same is an example of universality of the limiting statistical state at the hard edge;
see e.g.~\cite[\S 7.2.5]{Fo10}. In the Laguerre case, we have from \cite[Eq.~(2.14)]{FT19} that
\begin{multline}\label{3.7c}
K^{\rm h,L}_{\infty,1}(x,y) =
\alpha_1^{\rm L}(x,y) J_a(\sqrt{x})  J_a(\sqrt{y}) +
\beta_1^{\rm L}(x,y) (xy)^{1/2} J_a'(\sqrt{x})  J_a'(\sqrt{y}) \\
+ \gamma_1^{\rm L}(x,y) 
J_a(\sqrt{x}) \sqrt{y} J_a'(\sqrt{y}) +
\gamma_1^{\rm L}(y,x) 
J_a(\sqrt{y}) \sqrt{x} J_a'(\sqrt{x}),
\end{multline}
where $\alpha_1^{\rm L}(x,y),\beta_1^{\rm L}(x,y),
\gamma^{\rm L}_1(x,y)$ are low order symmetric polynomials in $x,y$ (explicity, equal to $-1/192$ times $a^2+x+y$, times 1, and times 2 respectively).
Explicit calculation shows that the Jacobi analogue of
\eqref{3.7c}, defined by replacing the superscript L by J throughout, with the left hand side then referring to the expansion \eqref{3.7b}, displays the same structure with low order symmetric polynomials of the very same degree.

\begin{prop}\label{P4.2}
With notation as defined above,
\begin{equation}\label{3.7d}
\alpha_1^{\rm J}(x,y) = {1 \over 48} ( - 2 a^2 - 3 b^2 + x + y), \quad \beta_1^{\rm J}(x,y) =  {1 \over 48}, \quad \gamma_1^{\rm J}(x,y) =  - {1 \over 48}.
\end{equation}
\end{prop}

\begin{proof}
This results from a computer algebra calculation (using Mathematica) starting with the functional form \eqref{3.2J}. As a detail, for the ${}_2 F_1$ function asymptotics, one isolates the general $k$-th term of the corresponding series, and then uses internal commands to expand for large $N_{\rm hJ}'$ before resumming (this strategy was suggested by Google Gemini AI). Another detail is to convert between ${}_0 F_1$ and Bessel functions, and to use a Bessel function identity to convert from $J_{a+1}(z)$ to $J_a'(z)$.
\end{proof}

In comparing the first correction with structure \eqref{3.7c} to the limiting functional form  \eqref{3.1}, a striking feature is that the denominator present in the finite $N$ expression \eqref{3.2J} has cancelled out. By analogy with the corresponding Gaussian unitary ensemble (GUE) and LUE expansions at the soft edge \cite{Bo24,YZ26}, this feature is expected of all higher order corrections in \eqref{3.1b} and \eqref{3.7b} too. It is further expected that the structural formula
\eqref{3.7c} remains valid at higher order, with polynomial coefficients that increase in degree for each new term. This is readily verified by explicit (computer algebra) computation for the second order correction.

Consider the expansion \eqref{3.7c} with each subscript 1 replaced by 2 (second correction term), and each superscript L replaced by J (Jacobi case). We have 
\begin{prop}\label{P4.3}
With notation as defined above,
\begin{align}\label{3.7d2}
\alpha_2^{\rm J}(x,y) &=  {1 \over 5760} \Big ( 8 a^4 - 2 a^2 + 21 (x^2 + y^2) + 
 16 x y + (x + y) (6 - 75 b^2 - 33 a^2) \Big ), \nonumber \\
 \beta_2^{\rm J}(x,y) &= {1 \over 2880} (-1 + 4 a^2 + 3( x +  y)), \nonumber \\
\gamma_2^{\rm J}(x,y) &=  {1 \over 11520}\Big (24 + 5 a^4 - 39 x - 29 y + a^2 (4 + 30 b^2 - 15 x + 5 y) \nonumber  \\ & \hspace*{6cm} + 
  5 (9 b^4 + x^2 - 6 b^2 (4 + x) + x y - y^2) \Big).  
\end{align}
\end{prop}

\begin{remark}\label{S4.3}(Special case of the classical groups.)
The mapping between the eigenvalue PDF for the classical groups and specialisations of the JUE is specified in \eqref{1.3d}. Examining \eqref{1.3d}, and adjusting $N$ so that on the right hand side the first entry is equal to $N$, one sees that the corresponding values of $N_{\rm hJ}'$ as specified by \eqref{3.1J} are $N-1/2,N,N,N+1/2$ respectively. This is in keeping with the known correlation kernels for the classical groups;
see \cite[Prop.~5.5.3]{Fo10}. For example
\begin{equation}\label{3.7e}
K_N^{{\rm O}^\pm(2N+1)}(\theta, \phi) =
{1 \over 2 \pi} \bigg (
{\sin N (\theta - \phi) \over \sin((\theta - \phi)/2)}
\mp {\sin N (\theta + \phi) \over \sin((\theta + \phi)/2)} \bigg ).
\end{equation}
For the mapping to the Jacobi ensemble we require a change of variables $\theta = \arccos(1 - 2 x)$,
$\phi = \arccos(1 - 2 y)$.
Using a hard edge scaling 
\begin{equation}\label{3.7e1}
\theta = X/N, \quad \phi = Y/N
\end{equation}
as is appropriate for angular variables, and considering the power series (in $\sqrt{\lambda}$) of $\arccos(1 - 2 \lambda)$,
we see that $X,Y$ permit series expansions in $\sqrt{x}/N, \sqrt{y}/N$, involving only powers of $N^{-2}$ and with $x= X^2, y = Y^2$ to leading order. Hence the fact that the kernels \eqref{3.7e} with the angular hard edge variables \eqref{3.7e1} admit a $1/N^2$ expansion is consistent with the expansion \eqref{3.7b} in the case $a=-b=\pm 1/2$. This example is further developed in Appendix A.
\end{remark}

\subsection{The density and spacing distributions}\label{S4.4}
The hard edge scaled density is the case $x=y$ of the hard edge correlation kernel. Its limiting form, $r_0^{\rm h}(y)$ say, as follows by taking the limit $x \to y$ in \eqref{3.1} 
and simplifying using Bessel function identities 
is \cite{Fo93}
\begin{equation}\label{3.7f}
r_0^{\rm h}(y) = {1 \over 4} \Big ( 
( (J_a(\sqrt{y}))^2 - J_{a+1}(\sqrt{y})
J_{a-1}(\sqrt{y}) \Big ) =
{1 \over 4} \Big ( \Big ( 1 - {a^2 \over y} \Big )
(J_a(\sqrt{y}))^2 +  (J_a'(\sqrt{y}))^2 \Big ).
\end{equation}
With $\rho_N^{\rm JUE}(x)$ the eigenvalue density of the JUE and $\rho_N^{\rm h,J}(y) := (1/4(N_{\rm hJ}')^2) \rho_N^{\rm JUE}(y/4(N_{\rm hJ}')^2)$ its hard edge scaling, we read off from
\eqref{3.7b} the expansion
\begin{equation}\label{3.7g}
\rho_N^{\rm h,J}(y) = r_0^{\rm h}(y) +
{1 \over (N_{\rm hJ}')^2} r_1^{\rm h,J}(y) + 
{1 \over (N_{\rm hJ}')^4} r_2^{\rm h,J}(y) + \cdots,
\end{equation}
where
\begin{equation}\label{3.7h}
r_1^{\rm h,J}(y)  =  K^{\rm h,J}_{\infty,1}(y,y), \quad
r_2^{\rm h,J}(y)  =  K^{\rm h,J}_{\infty,2}(y,y),
\end{equation}
with $K^{\rm h,J}_{\infty,1}(y,y)$, $K^{\rm h,J}_{\infty,2}(y,y)$ in turn as implied by
\eqref{3.7c} and Propositions \ref{P4.2} and \ref{P4.3}. Explicitly, the first two corrections are
\begin{equation}\label{3.8J}
r_1^{\rm h,J}(y)={1\over48}\Big((2y-2a^2-3b^2)(J_a(\sqrt{y}))^2
-2\sqrt{y}J_a(\sqrt{y})J_a'(\sqrt{y})
+y(J_a'(\sqrt{y}))^2\Big),
\end{equation}
and
\begin{multline}\label{3.9J}
r_2^{\rm h,J}(y)={}
{a^2(4a^2-1)+(6-33a^2-75b^2)y+29y^2\over2880}
(J_a(\sqrt{y}))^2  \\
+{5(y-a^2-3b^2)^2+4a^2-120b^2-68y+24\over5760}
\sqrt{y}J_a(\sqrt{y})J_a'(\sqrt{y})
+{4a^2-1+6y\over2880}\,y(J_a'(\sqrt{y}))^2.
\end{multline}

Let $K_N^{\rm h,J}(x,y)$ denote the scaled kernel in \eqref{3.5J}, and define the conditioned gap probability generating function
\begin{equation}\label{3.24J}
\mathcal E^{\rm h,J}((0,y);\xi,N)
=\sum_{k=0}^N(1-\xi)^kE_{N}^{\rm JUE}(k;(0,y/(4(N_{\rm hJ}')^2)))  
 =\det(I-\xi \mathbb K_N^{\rm h,J})_{L^2(0,y)},
\end{equation}
where in the last equality, the $\det(\cdot)$ refers to the Fredholm determinant, with $\mathbb K_N^{\rm h,J}$ representing the integral operator on $(0,y)$ with kernel $K_N^{\rm h,J}(x,y)$. Here $E_{N}^{\rm JUE}(k;(0,s))$ is the probability of exactly $k$ eigenvalues in the interval $(0,s)$ for the JUE.
For $\xi=1$, $\mathcal E^{\rm h,J}((0,y);\xi,N)$ reduces to the gap probability of no eigenvalues in the indicated interval as seen from the $k=0$ term in the sum.
Introduce the scaled $k$-point correlation function
\begin{align}\label{kc}
\rho_{(k),N}^{\rm h,J}(x_1,\ldots,x_k)
&={1\over(4(N_{\rm hJ}')^2)^k}
\rho_{(k),N}^{\rm JUE}\left({x_1\over4(N_{\rm hJ}')^2},
\ldots,{x_k\over4(N_{\rm hJ}')^2}\right) \nonumber \\
&=\det\left[K_N^{\rm h,J}(x_i,x_j)\right]_{i,j=1}^k,
\qquad 1\leq k\leq N.
\end{align}
In particular, $\rho_{(1),N}^{\rm h,J}(x)=\rho_N^{\rm h,J}(x)$ gives the scaled density. Then the generating function \eqref{3.24J} has an alternative representation in terms of the scaled correlation functions as
\begin{equation}\label{gen-corr}
\mathcal E^{\rm h,J}((0,y);\xi,N)
=1+\sum_{k=1}^N{(-\xi)^k\over k!}
\int_{(0,y)^k}\rho_{(k),N}^{\rm h,J}(x_1,\ldots,x_k)
\,dx_1\cdots dx_k.
\end{equation}
The kernel expansion \eqref{3.7b} implies a large $N'_{\rm hJ}$ expansion of the correlation functions in even powers of $1/N_{\rm hJ}'$,
{\small $$ 
\rho_{(k),N}^{\rm h,J}(x_1,\ldots,x_k) = {}
r_{(k),0}^{\rm h,J}(x_1,\ldots,x_k)
+\frac{1}{(N_{\rm hJ}')^2}r_{(k),1}^{\rm h,J}(x_1,\ldots,x_k)
+\frac{1}{(N_{\rm hJ}')^4}r_{(k),2}^{\rm h,J}(x_1,\ldots,x_k)+\cdots,
$$
}%
where the leading term is universal (here meaning the same for the Laguerre and Jacobi hard edge scalings).
This combined with \eqref{gen-corr} implies, subject to assumptions on the nature of the error term not addressed here
\cite[Lemma 2.1]{BFM17}, an expansion of the same structure
\begin{align}\label{3.26J}
\mathcal E^{\rm h,J}((0,y);\xi,N) =
\mathcal E^{0,\rm h}((0,y);\xi)
+\frac{1}{(N_{\rm hJ}')^2}\mathcal E^{1,\rm h,J}((0,y);\xi) +\frac{1}{(N_{\rm hJ}')^4}\mathcal E^{2,\rm h,J}((0,y);\xi)+\cdots.
\end{align}

There have been
significant recent developments in relation to scaled large $N$ generating function hard edge expansions of the type \eqref{3.26J}.
The point is that there are differential relations linking the corrections $\mathcal E_i$ ($i=1,2,\dots)$ to the limit law $\mathcal E_0$, which most significantly are independent of $\xi$.
These were initiated and implemented to high order in \cite{Bo25a} in the context of the Gaussian and Laguerre soft edge,
and  then shown to hold true for the leading order in the circular ensemble bulk \cite{FS26}, and Laguerre ensemble hard edge \cite{FRS26}.
As observed in \cite{Bo25}, these being independent of $\xi$
imply analogous differential relations 
for the scaled large $N$ density expansion linking the corrections to the limit law.
Moreover, it was demonstrated in \cite{Bo25} that this line of reasoning can be reversed: thus knowing the differential relations for the density corrections can be used to deduce the same for the conditioned gap probability generating function. Specifically, for ${\mathcal P}_j^{\rm J}$, ${\mathcal P}^{*,\rm J}_j$
differential operators of degree $2j-1$, 
\begin{align}\label{3.27}
r_j^{\rm h,J}(y) = {d \over dy}  {\mathcal P}_j^{\rm J} r_0^{\rm h}(y) \quad \implies \quad 
\mathcal E^{j,\rm h,J}((0,y);\xi) =   {\mathcal P}^{*,\rm J}_j
{d \over dy} \mathcal E^{0,\rm h}((0,y);\xi)
\end{align}
with further structural properties of
$\{ r_j^{\rm h,J}(y) \}$ providing a systemic way to compute ${\mathcal P}^{*, \rm J}_j$ directly (see \S \ref{S4.5} in relation to the case $j=2$).
Moreover, in the simplest case $j=1$, $
{\mathcal P}^{*,\rm J}_j = {\mathcal P}_j^{\rm J}
$;  for $j \ge 2$ $\mathcal P_j$ is not unique as noted in Remark \ref{R4.density-ambiguity}, point 1., below.

From knowledge of the explicit form of $r_1^{\rm h,J}(y)$ \eqref{3.8J}, we can readily calculate ${\mathcal P}_1^{\rm J}$
and ${\mathcal P}^{*,\rm J}_1$.

\begin{prop}\label{P4.4}
 Let $\mathcal D_y = {d \over d y} y$.  We have
 \begin{equation}\label{fg1}  
\Big ( - {1 \over 6} \mathcal D_y^2 + {1 \over 12} \mathcal D_y y
+ {1 \over 6} \Big ( 1 - {1 \over 2} (a^2 + 3 b^2) \Big )\mathcal D_y \Big )
r_0^{\rm h}(y) = r_1^{\rm h,J}(y), 
 \end{equation}
 Consequently (under the assumption leading to \eqref{3.27} in the case $j=1$)
 \begin{equation}\label{4.28}
 \mathcal E^{1,\rm h,J}((0,y);\xi) =
 {\mathcal P}^{*,\rm J}_1 {d \over dy} \mathcal E^{0,\rm h}((0,y);\xi), \quad  {\mathcal P}^{*,\rm J}_1 =
 \Big ( - {1 \over 6} y^2 {d \over d y} 
+ {1 \over 12}  (y - (a^2 + 3 b^2))  y \Big ).
 \end{equation}
\end{prop}

\begin{proof}
The first of these follows from the explicit formulas for $ \mathcal D_y^2 r_0(y)$
and $ \mathcal D_y r_0(y)$ presented in \cite[Eq,~(4.40)]{FRS26},
and noting too the operator identity $\mathcal D_y y = y + y \mathcal D_y$. The second follows from \eqref{3.27} with
$
{\mathcal P}^{*,\rm J}_1 = {\mathcal P}_1^{\rm J}
$.
\end{proof}

\begin{remark}\label{R4.density-ambiguity} ${}$
1.~For $j \ge 2$, the operators $\mathcal P_j^{\rm J}$ in
\eqref{3.27} are not unique, due to the possibility of the
replacement
\begin{equation}\label{3.23J}
\mathcal P_j^{\rm J}\mapsto
\mathcal P_j^{\rm J}+\lambda \mathcal D_0^{\rm JUE,h}.
\end{equation}
for some constant $\lambda$ (which may itself be a differential operator for $j>2$) and third order differential operator
$\mathcal D_0^{\rm JUE,h}$.
The main point here is that
there is a known differential operator such that $\mathcal D_0^{\rm JUE,h}r_0^{\rm h}=0$ \cite{RF19} (see 
\eqref{3.12J} below for  its explicit form).\\
2.~There is the simple exact result \cite[Eq.~(1.11) with 
$\beta = 2$, $\lambda_2=0$, $\lambda_1=b$, $s \mapsto (1-y/(2N_{\rm h J}')^2)$]{FK23}, 
\begin{equation}\label{3.23Ja}
\mathcal E_N^{\rm h,J}((0,y);\xi,N) \bigg |_{\substack{\xi=1 \\ a=0}} =
\Big ( 1 - y/(2N_{\rm h J}')^2 \Big )^{(N_{\rm h J}')^2 - (b/2)^2}.
\end{equation}
Hence with $\xi =1$, $a=0$,
\begin{equation}\label{3.23Jb}
\mathcal E^{0,\rm h}((0,y);\xi) = e^{-y/4}, \quad
\mathcal E^{1,\rm h,J}((0,y);\xi) = e^{-y/4} \Big (
 {y b^2 \over 16} - {y^2 \over 32} \Big ),
\end{equation} 
which indeed are related by the differential relation \eqref{4.28}. \\
3.~In the case of the LUE, it has been shown in 
\cite[Appendix C]{FRS26} that a proof of the analogue of \eqref{4.28} can be given based on a $\sigma$-Painlev\'e characterisation of the  
finite $N$ generating function. An alternative method, applicable too in the proof of differential relations beyond 
the first correction is to use instead the Fredholm determinant expression \eqref{3.24J} as the starting point
\cite{Bo24}. Although not pursued in the present work, it is expected that either method can be used to give a proof of
\eqref{4.28}. Specifically, for the $\sigma$-Painlev\'e method, it is known from \cite{FW04} that the  
finite $N$ generating function permits an evaluation in terms of a $\sigma$-PVI transcendent (see also \cite[\S 8.3.1]{Fo10}).
\end{remark}

\subsection{Coupled differential equations for the density and related differential operators}\label{S4.5}
It is known that the finite $N$ JUE density $\rho_N^{\rm JUE}$ satisfies a third-order linear differential
equation \cite[Theorem~2.1, Eqs.~(2.19)--(2.20)]{RF19}. Under the optimal hard edge scaling, the differential operator is a polynomial in
$1/(N_{\rm hJ}')^2$.

\begin{prop}\label{P4.4a}
Write $\partial_y={d\over dy}$. For $N\geq1$, $a,b>-1$ and $0<y<4(N_{\rm hJ}')^2$, the scaled density obeys
\begin{equation}\label{3.11J}
\left(\mathcal D_0^{\rm JUE,h}+\frac{1}{(N_{\rm hJ}')^2}\mathcal D_1^{\rm JUE,h}
+\frac{1}{(N_{\rm hJ}')^4}\mathcal D_2^{\rm JUE,h}
+\frac{1}{(N_{\rm hJ}')^6}\mathcal D_3^{\rm JUE,h}\right)\rho_N^{\rm h,J}(y)=0,
\end{equation}
where
{ \small \begin{align}\label{3.12J}
& \mathcal D_0^{\rm JUE,h}= y^3\partial_y^3+4y^2\partial_y^2
+y(y-a^2+2)\partial_y+{y\over2}-a^2, \nonumber \\
& \mathcal D_1^{\rm JUE,h}= -{3y^4\over4}\partial_y^3-4y^3\partial_y^2
+{y^2\over4}(2a^2-b^2-2y-16)\partial_y 
+{y\over8}(5a^2-b^2-3y-4), \nonumber \\
&\mathcal D_2^{\rm JUE,h}= {3y^5\over16}\partial_y^3
+{5y^4\over4}\partial_y^2
+{y^3\over16}(-a^2+b^2+y+28)\partial_y 
 +{y^2\over32}(-3a^2+3b^2+2y+12), \nonumber \\
& \mathcal D_3^{\rm JUE,h}= -{y^6\over64}\partial_y^3
-{y^5\over8}\partial_y^2-{7y^4\over32}\partial_y-{y^3\over16}.
\end{align}}
Consequently, the coefficients $r_j^{\rm h}(y)$ in the hard edge expansion \eqref{3.7g} satisfy, for $j\geq0$,
\begin{equation}\label{3.13J}
\mathcal D_0^{\rm JUE,h}r_j^{\rm h,J}
=-\sum_{l=1}^{3}\mathcal D_l^{\rm JUE,h}r_{j-l}^{\rm h,J},
\end{equation}
where $r_k^{\rm h,J}=0$ for $k<0$ and $r_0^{\rm h,J}:=r_0^{\rm h}$.
\end{prop}

\begin{proof}
Substitute
$t=y/(2N_{\rm hJ}')^2$ and $2N+a+b=2N_{\rm hJ}'$ into the density equation of \cite[Eq.~(2.19)]{RF19}.
Using $d/dt=(2N_{\rm hJ}')^2\partial_y$ and collecting powers of
$(N_{\rm hJ}')^{-2}$ gives \eqref{3.11J}--\eqref{3.12J}.
Equating powers of $(N_{\rm hJ}')^{-2}$ after inserting the expansion \eqref{3.7g} then gives \eqref{3.13J}.
\end{proof}

In particular, the equations for the first two corrections are
\begin{align}\label{3.14J}
\mathcal D_0^{\rm JUE,h}r_0^{\rm h} =0, \quad
\mathcal D_0^{\rm JUE,h}r_1^{\rm h,J} =-\mathcal D_1^{\rm JUE,h}r_0^{\rm h}, \quad 
\mathcal D_0^{\rm JUE,h}r_2^{\rm h,J} =-\mathcal D_1^{\rm JUE,h}r_1^{\rm h,J}
-\mathcal D_2^{\rm JUE,h}r_0^{\rm h} 
\end{align}
(compare the second of these to \eqref{4.28}). With the functional form of $r_0^{\rm h}$ known from \eqref{3.7f}, these provide inhomogeneous differential equations for 
$r_1^{\rm h,J}$ and $r_2^{\rm h,J}$.
The equations are supplemented by the boundary conditions
\begin{equation}\label{3.15J}
r_j^{\rm h,J}(y)={g_j\over4^{a+1}\Gamma(a+1)\Gamma(a+2)}\,y^a
+O_{a,b}(y^{a+1}),\qquad y\downarrow0,\quad j=0,1,2,
\end{equation}
where
\begin{align}\label{3.16J}
g_0={}&1,\qquad g_1=-{(a+1)(a^2+2a+3b^2)\over12}, \nonumber \\
g_2={}&{a(a+1)\over1440}
\Big(5a^4+16a^3+30a^2b^2+4a^2-4a
+45b^4-120b^2+24\Big).
\end{align}
Indeed, the exact finite $N$ Christoffel--Darboux formula gives
\begin{equation}\label{3.17J}
\lim_{y\downarrow0}y^{-a}\rho_N^{\rm h,J}(y)
={\Gamma(N+a+1)\Gamma(N+a+b+1)\over
 (4(N_{\rm hJ}')^2)^{a+1}\Gamma(N)\Gamma(N+b)
 \Gamma(a+1)\Gamma(a+2)}.
\end{equation}
Expanding the gamma ratios gives \eqref{3.16J}, consistent with
\eqref{3.7f}, \eqref{3.8J} and \eqref{3.9J}.
One notes in particular that the first of the equations in 
\eqref{3.14J}, which is the homogeneous part of the following two, has a unique solution with this small $y$ behaviour. This is consistent with the allowed solution of the homogeneous part always being $A r_0^{\rm h}(y)$ for some constant $A$. In fact, as found 
for the expansion analogous to \eqref{3.7g} for the LUE \cite{FRS26},
there is no contribution from the homogeneous part in the solutions of the second and third equations of \eqref{3.14J}.

Let
\begin{equation}\label{4.A}
b_1(y) = {1 \over y} (J_a(\sqrt{y}))^2, \quad
b_2(y) =  (J_a'(\sqrt{y}))^2, \quad
b_3(y) = {1 \over \sqrt{y}} J_a(\sqrt{y} )J_a'(\sqrt{y}).
\end{equation}
Generally, the solutions of coupled equations \eqref{3.13J} have the structure 
\begin{equation}\label{2.RL}
 r_j^{\rm h,J}(y) = \alpha_j^{\rm h,J}(y) b_1(y) +
  \beta_j^{\rm h,J}(y)  b_2(y)  + \gamma_j^{\rm h,J}(y)
  b_3(y),
\end{equation} 
where $ \alpha_j^{\rm h,J}(y), \beta_j^{\rm h,J}(y),   \gamma_j^{\rm h,J}(y)$ are polynomials, with order increasing as $j$ increases. This follows from the fact that $\mathcal D_y b_i(y)$, with $\mathcal D_y$ as in Proposition \ref{P4.4}, have the structure of the right hand side of \eqref{2.RL}, with the polynomial coefficients of degree no bigger than one, and consideration of the boundary condition. The same structure holds for the LUE \cite{FRS26,LMZ26}. One sees that it is exhibited by \eqref{3.8J} and \eqref{3.9J}, as well as by the limiting density \eqref{3.7f}.

In an analogous problem (soft edge scaling of the Laguerre unitary ensemble) it has been shown in 
\cite[\S 5]{Bo25} how the structure \eqref{2.RL} can be used to provide a computational scheme for the operators
$\partial_y \mathcal P_j^{*,\rm J}$,
where $ \mathcal P_j^{*,\rm J} = \sum_{k=0}^{2j-1} c_k(y) \partial^k_y$, appearing in
\eqref{3.27}.
The method is essentially the same as that used for the proof of Proposition \ref{P4.4a}, with the refinement of making use of the fact  that the functions (now involving products of Airy functions and its derivative) 
playing the role of \eqref{4.A} are algebraically independent, and with the assumption that the unknown polynomials
$c_k(y)$  are to be determined in descending order $k=2j-1,\dots, 0$. The algebraic independence of the Bessel function and its derivative, for orders $a$ in \eqref{4.A} not equal to a half integer, is well known (it follows by the argument revised in the opening paragraph of \cite{Bo26a}). This allows us to proceed via (a modification of) the method of \cite{Bo25} to determine $\mathcal P_j^{*,\rm J}$ for $j=2$

\begin{prop}
We have 
\begin{multline}\label{3.19Ja}
\mathcal P_2^{*\rm J}= {y^4\over72}\partial_y^3
+{y^3\over360}\Big(5(a^2+3b^2-y)+8\Big)\partial_y^2
+{y^2\over1440}\Big(5(y-a^2-3b^2)^2
 \\
+4a^2-60b^2-48y+16\Big)
\partial_y 
+{y\over360}\Big(4y^2+(1-3a^2-15b^2)y
+a^2(6-a^2-15b^2)\Big).
\end{multline}
\end{prop}

\begin{proof}
As a minor modification of the beginning strategy of \cite{Bo25}, we seek to express $P_2^{*\rm J}$ in the form
$y \mathcal( \sum_{k=1}^4 c_k(y) \mathcal D_y^{k-1})$, with each $c_k(y)$ a polynomial of degree $4-k$. Then from 
\eqref{3.27} we have
 \begin{equation}\label{H1s}
r_2^{\rm h,J}(y)= 
 \sum_{k=1}^4 \Big ( y c_{k}'(y) +
c_k(y) \mathcal D_y \Big ) \mathcal D_y^{k-1} r_0^{\rm h}(y).
\end{equation}
We know the explicit form of the polynomial coefficients 
of the $b_i(y)$ in \eqref{2.RL} for $r_2^{\rm h,J}(y)$ from
\eqref{3.9J}. We want to compute the same for the right hand side of \eqref{H1s}.
For this task, in addition to the expansion of 
$ \mathcal D_y^2 r_0(y)$
and $ \mathcal D_y r_0(y)$ 
in the form of the right hand side of \eqref{2.RL} as
presented in \cite[Eq.~(4.40)]{FRS26}, we require the same
for $ \mathcal D_y^j r_0(y)$, $j=3,4$. Use of computer algebra shows
{\small
\begin{align*}
 \mathcal D_y^3 r_0(y) & = {1 \over 8} \Big ( (2+a^2-y)(J_a(\sqrt{y})^2 + 4 \sqrt{y} J_a(\sqrt{y}) J_a'(\sqrt{y}) + y (J_a'(\sqrt{y}))^2 \Big ), \\
 \mathcal D_y^4 r_0(y) & = {1 \over 8} \Big ( (2+3a^2-4y)(J_a(\sqrt{y})^2 + 2(4+a^2-y) \sqrt{y} J_a(\sqrt{y}) J_a'(\sqrt{y}) + 3y (J_a'(\sqrt{y})^2 \Big ).
\end{align*}}%
With these expansions, and the polynomials $c_k(y)$ written in terms of their coefficients, the right hand side of \eqref{H1s}
can now be expressed in terms of the $b_i(y)$. Equating the polynomial coefficients on both sides gives a system of linear equations for
the coefficients specifying the $c_k(y)$. In keeping with
Remark \ref{R4.density-ambiguity}, this linear system has  a one parameter family of solutions depending on $c_4(y)=:c_4$, which is a constant. 

To fix $c_4$, one can consider the leading 
small $y$ form of  \eqref{gen-corr} at order $\xi^2$.
This is given by a double integral over $\rho_{(2),N}^{\rm h,J}(t_1, t_2)$. From its definition, for $t_1,t_2\to 0$ the latter can be expanded
\begin{align}\label{3.twopoint-endpointJ}
\rho_{(2),N}^{\rm JUE}(t_1,t_2)
&=B_N(t_1t_2)^a(t_1-t_2)^2
\big(1+{\rm O}(t_1+t_2)\big), \quad
B_N=N(N-1){Z_{N-2}(a+4,b)\over Z_N(a,b)}.
\end{align}
Here $Z_N(a,b)$ is given by a particular Selberg integral (see e.g.~\cite[Ch.~4]{Fo10})
\begin{align}\label{4.43a}
    Z_N(a,b)
=\int_{(0,1)^N}\prod_{l=1}^N t_l^a(1-t_l)^b
\prod_{i<l}(t_i-t_l)^2\,dt_1\cdots dt_N=\prod_{l=0}^{N-1}
{\Gamma(a+l+1)\Gamma(b+l+1)\Gamma(l+2)\over
\Gamma(a+b+N+l+1)}.
\end{align}
Using
\begin{align*}
    {1\over2}\int_{(0,s)^2}(x_1x_2)^a(x_1-x_2)^2\,dx_1dx_2
={s^{2a+4}\over(a+1)(a+2)^2(a+3)},
\end{align*}
and making the hard edge change of variables shows
\begin{align}\label{3.twopoint-finiteJ}
[\xi^2]\mathcal E^{\rm h,J}((0,y);\xi,N)
=C_Ny^{2a+4}+{\rm O}(y^{2a+5}), \quad
C_N={B_N\over
(4(N_{\rm hJ}')^2)^{2a+4}(a+1)(a+2)^2(a+3)}.
\end{align}
Denoting the right hand side of \eqref{3.17J} by $k_N$, we see that after substituting the Selberg product and using the gamma recurrence, $C_N$ can be rewritten as
\begin{align}
    C_N={k_N^2\over16(a+1)^2(a+2)^4(a+3)^2}
\left(1-{(a+2)^2+b^2\over2(N_{\rm hJ}')^2}
+{\big((a+2)^2-b^2\big)^2\over16(N_{\rm hJ}')^4}\right).
\end{align}
In keeping with \eqref{3.15J}, and with $g_1,g_2$ given in \eqref{3.16J}, $k_N$ has the  expansion
\begin{align*}
    k_N=c_a\left(1+{g_1\over(N_{\rm hJ}')^2}
+{g_2\over(N_{\rm hJ}')^4}+{\rm O}((N_{\rm hJ}')^{-6})\right),\quad c_a={1\over4^{a+1}\Gamma(a+1)\Gamma(a+2)}.
\end{align*}
As a result, we have
\begin{align}
    C_N=C_a\left(1+{d_1\over(N_{\rm hJ}')^2}
+{d_2\over(N_{\rm hJ}')^4}+{\rm O}((N_{\rm hJ}')^{-6})\right),
\end{align}
where
\begin{align}\label{3.twopoint-coeffJ}
&C_a ={c_a^2\over16(a+1)^2(a+2)^4(a+3)^2}, \quad 
d_1 =2g_1-{(a+2)^2+b^2\over2}, \nonumber \\
&d_2 =g_1^2+2g_2-\big((a+2)^2+b^2\big)g_1
+{\big((a+2)^2-b^2\big)^2\over16}.
\end{align}
Substituting these into \eqref{3.twopoint-finiteJ}, we get the edge limit conditions
{\small \begin{align}\label{2point-nor}
    [\xi^2]\mathcal E_0^{\rm h,J}((0,y);\xi)
=C_ay^{2a+4}+{\rm O}(y^{2a+5}), \quad
[\xi^2]\mathcal E_j^{\rm h,J}((0,y);\xi)
=C_ad_jy^{2a+4}+{\rm O}(y^{2a+5}),\: (j=1,2).
\end{align}}%
Applying $\mathcal P_2^{*\rm J}\partial_y$ to $y^{2a+4}$, and comparing the resulting leading term with the one in the second equation of \eqref{2point-nor} fixes $c_4$, and \eqref{3.19Ja} follows.
\end{proof}

\begin{remark} ${}$ \\
1.~The exact result \eqref{3.23Ja} gives that with $\xi = 1, a=0$
 \begin{equation}\label{dp1}
\mathcal E^{\rm 2,h,J}((0,y);\xi) = {e^{-y/4}y^2 \over 6144} \Big (
48 b^2 + 12 b^4 - 32 y - 12 b^2 y + 3 y^2 \Big ).
 \end{equation}
 This is consistent with the second equation in \eqref{3.27} 
 with $j=2$ and $\mathcal P_2^{*J}$ given by \eqref{3.19Ja}. \\
 2.~Combining the first and second equations in \eqref{3.14J} gives $\mathcal D_0^{\rm JUE,h} r_1^{\rm h,J} = -
 (\mathcal D_1^{\rm JUE,h} + c \mathcal D_0^{\rm JUE,h}) r_0^{\rm h,J}$ for any scalar (or in fact differential operator) $c$. Choosing $c = 3y/4$ then shows 
  \begin{equation}\label{dpt}
 \mathcal D_0^{\rm JUE,h} r_1^{\rm h,J} = -\tilde{D}_1^{\rm JUE,h} r_0^{\rm h,J}, \quad  \tilde{D}_1^{\rm JUE,h} :=
 -y^3 \partial^2_y + \frac{y^2}{4}(y - a^2 - b^2 - 10)\partial_y - \frac{y}{8}(a^2 + b^2 + 4).
  \end{equation}
  Note that $\tilde{D}_1^{\rm JUE,h}$ is a second order differential operator, whereas ${D}_1^{\rm JUE,h}$  is third order.
  \end{remark}

 \section{Optimal Jacobi hard edge scaled large $N$ expansions beyond the JUE}
 \subsection{Even $\beta$}
We first introduce the $k$-point correlation 
$\rho_{(k),\beta,N}(x_1,\dots,x_k)$ 
for the Jacobi $\beta$ ensemble. For this we require the Selberg integral (see e.g.~\cite[\S 4.1]{Fo10})
\begin{align}
S_N(\lambda_1,\lambda_2,\lambda) &:=
\int_0^1 dt_1 \cdots \int_0^1 dt_N \,
 \prod_{l=1}^N t_l^{\lambda_1}(1-t_l)^{\lambda_2}
 \prod_{1 \le j < k \le N} |t_k - t_j|^{2 \lambda} \nonumber \\
 &  = \prod_{j=0}^{N-1} {\Gamma (\lambda_1 + 1 + j\lambda)
\Gamma (\lambda_2 + 1 + j\lambda)\Gamma(1+(j+1)\lambda) \over
\Gamma (\lambda_1 + \lambda_2 + 2 + (N + j-1)\lambda) \Gamma (1 + \lambda )},
\end{align}
(cf.~\eqref{4.43a})
and the weight \eqref{WJ}. Recalling that the eigenvalue PDF for the Jacobi $\beta$ ensemble is proportional to \eqref{1.1J}, 
after modifying the exponents to be consistent with
\eqref{WJ} we see that the $k$-point correlation function, defined as the integral over all but $k$ of the eigenvalues in the eigenvalue PDF, is specified by
\begin{multline}
\rho_{(k),\beta,N}(x_1,\dots,x_k) = {N! \over (N - k)!}
{\prod_{l=1}^k w^{\rm J}(x_l) \over S_N(a,b,\beta/2)} \\
\times 
\int_0^1 dx_{k+1} \cdots \int_0^1 dx_N \, \prod_{l=k+1}^N
w^{\rm J}(x_l) \prod_{1 \le j < k \le N} |x_k - x_j|^\beta.
\end{multline}
In keeping with \eqref{kc} we define the corresponding hard edge scaled correlations
\begin{equation}\label{kc1}
\rho_{(k),N,\beta}^{\rm h,J}(x_1,\dots,x_k) :=
\Big ( {1 \over 4 (N_{\rm hJ'})^2} \Big )^k
\rho_{(k),N,\beta}
\Big ( {x_1 \over 4 (N_{\rm hJ'})^2},
\dots, {x_k \over 4 (N_{\rm hJ}')^2} \Big ).
\end{equation}
 
 We have seen that for $\beta =2$, where there is a determinantal structure, the correlation kernel determining the general $k$-point correlation is given by a certain ${}_2 F_1$ hypergeometric polynomial. While the determinantal structure is particular to $\beta = 2$, there is a polynomial structure present for all even $\beta$. However, now the polynomials are multivariable. They relate to the ${}_2 F_1$ hypergeometric polynomial too. In the definition of the latter, Jack polynomials $P_\kappa^{(\alpha)}(\mathbf x)$ replace the monomials $x^k$. One recalls (see e.g.~\cite[Ch.~12]{Fo10}) that Jack polynomials depend on a parameter $\alpha>0$, and are labelled by a partition $\kappa=(\kappa_1,\dots,\kappa_n)$, $\kappa_1 \ge \kappa_2 \ge \cdots \ge \kappa_n \ge 0$. These multivariable polynomials, dependent on $\mathbf x = (x_1,\dots,x_n)$, are homogeneous of degree $|\kappa| := \sum_{l=1}^n \kappa_l$. The rising Pochhammer symbols $(u)_k$ in the definition of ${}_2 F_1$ are replaced by
  \begin{equation}\label{4.3m}
\quad[u]_\kappa^{(\alpha)} := \prod_{l=1}^n {\Gamma(u - (l-1)/\alpha + \kappa_l) \over \Gamma(u - (l - 1)/\alpha)};
\end{equation}
see e.g.~\cite[Eq.~(12.46)]{Fo10}. The family of generalised hypergeometric functions of interest are then specified by
\begin{equation}\label{3.40}
  {\vphantom{F}}_2^{\mathstrut} F_1^{(\alpha)}(a_1,a_2;b_1; \mathbf x):=\sum_\kappa  {\alpha^{| \kappa |} \over h_\kappa'(\alpha)} 
 \frac{[a_1]^{(\alpha)}_\kappa
 [a_2]^{(\alpha)}_\kappa}{[b_1]^{(\alpha)}_\kappa} 
P_\kappa^{(\alpha)}(\mathbf x), 
\end{equation}
for a certain $h_\kappa'(\alpha)$; see e.g.~\cite[\S 13.1]{Fo10}. For $a_1$ a negative integer, \eqref{3.40} specifies a (multivariable) polynomial. A result of \cite[\S 13.2.4]{Fo10} gives that for $\beta$ even
\begin{eqnarray}\label{15.ge.1}
&& \rho_{(k),\beta,N}(x_1,\dots,x_k)    =  {N! \over (N-k)!} {S_{N-k}(a+k\beta,b,\beta/2) \over
S_{N}(a,b,\beta/2)} \prod_{l=1}^k w^{\rm J}(x_l)
\prod_{j<l}^k | x_l - x_j|^\beta
  \nonumber \\
&&\quad \times 
{\vphantom{F}}_2^{\mathstrut}  F_1^{(\beta/2)}(-N+k,(2/\beta)(a+b+1) + k  + N - 1;
2 a /\beta + 2k;t_1,\dots, t_{\beta k}) \Big |_{\{t_j\} \mapsto \{x_j\}}
\end{eqnarray}
where
\begin{equation}\label{15.tr}
 \quad t_l = x_j \quad {\rm for} \quad l=1+(j-1)\beta, \dots, j \beta
\: \: (j=1,\dots,k).
\end{equation}
Notice that in the notation of \eqref{3.40}, the generalised hypergeometric function in \eqref{15.ge.1} has $a_1 = -N+k$ ---
a negative integer --- and so corresponds to a multivariable polynomial.

\begin{prop} \label{prop5.1}
Let $\beta$ be even. With $\mathbf x :=(x_1,\dots,x_k)$
set
\begin{equation}\label{Trs}
W_k(\mathbf x) := \prod_{l=1}^k x_l^a \prod_{ j   < l}^k
|x_l - x_j|^\beta, \quad c_{k,\beta} := \Gamma(1+\beta/2)^k  \prod_{j=0}^{2k-1} \frac{1}{\Gamma(a + 1 + \beta j/2)}.
\end{equation}
The large $N_{\rm hJ}'$ expansion of
\eqref{kc1} is an asymptotic expansion in 
powers of $(N_{\rm hJ}')^{-2}$. The leading term is independent of
$(N_{\rm hJ}')^{-2}$ and given by
\begin{equation}\label{Trs1}
\lim_{N \to \infty} \rho_{(k),N,\beta}^{\rm h,J}(\mathbf x)
= 4^{-k} (\beta/2)^{2k(a+1)+\beta k(k-1)-k}c_{k,\beta} W_k(\mathbf x/4) 
{\vphantom{F}}_0^{\mathstrut}  F_1^{(\beta/2)}(2a/\beta+2k;- \mathbf x/4).
\end{equation}
\end{prop}

\begin{proof}
Introducing $N_{\rm hJ}'$ as specified in \eqref{3.1J}, we see that the first two arguments of 
${\vphantom{F}}_2^{\mathstrut}  F_1^{(\beta/2)}$ in  
\eqref{15.ge.1} are
$$
-N_{\rm hJ}'-{1 \over 2} + {a + b +1 \over \beta} + k, \quad
N_{\rm hJ}'-{1 \over 2} + {a + b +1 \over \beta} + k.
$$
Due to the fact from \eqref{3.40}
that 
${\vphantom{F}}_2^{\mathstrut}  F_1^{(\beta/2)}(a_1,a_2;b_1;\mathbf x)$ is symmetric in $a_1, a_2$, we see immediately that this factor in \eqref{3.40} is even in
$N_{\rm hJ}'$. Moreover, applying an analysis presented in detail in
\cite[\S 3.2]{FS26a}, we can readily establish that the scaling of this generalised hypergeometric function as given in \eqref{kc1}
is analytic in $1/N_{\rm hJ}'$, for any fixed polydisk of $\mathbf x$. In particular, its large $N_{\rm hJ}'$ limit is given by
\begin{equation}\label{5.9o}
{\vphantom{F}}_0^{\mathstrut}  F_1^{(\beta/2)}(2a/\beta+2k;- \mathbf x/4),
\end{equation}
defined as for \eqref{3.40} but with the terms dependent of $a_1,a_2$ removed.

For the ratio of Selberg integrals in \eqref{15.ge.1} we have
\begin{multline}
\frac{S_{N-k}(a+2k\lambda,b,\lambda)}{S_N(a,b,\lambda)} =  \Gamma(1+\lambda)^k  \prod_{j=0}^{2k-1} \frac{1}{\Gamma(a + 1 + j\lambda)}\\  
\times \prod_{j=0}^{k-1} \frac{\Gamma(a + 1 + (N+j)\lambda) \, \Gamma(a + b + 2 + (N + j - 1)\lambda)}{\Gamma(b + 1 + (N -  j-1)\lambda) \, \Gamma(1 + (N - j)\lambda)},  
\end{multline}
where we have set $\lambda := \beta/2$ for convenience. Now multiplying by $N!/(N-k)!$, using the gamma function recurrence in the final factor of the denominator to effectively cancel this term, and introducing \eqref{3.1J}, we obtain
\begin{equation}\label{5.9}
    \lambda^{-k}\Gamma(1+\lambda)^k  \prod_{j=0}^{2k-1} \frac{1}{\Gamma(a + 1 + j\lambda)}
    \prod_{j=0}^{k-1} {\Gamma(\lambda N_{\rm hJ}' + c_j +1) \over \Gamma(\lambda N_{\rm hJ}' - c_j)}
    {\Gamma(\lambda N_{\rm hJ}' + d_j +1) \over \Gamma(\lambda N_{\rm hJ}' - d_j)},
    \end{equation}
where $c_j = \lambda(j+{1 \over 2}) + {1 \over 2} (a - b - 1)$,
$d_j = \lambda(j-{1 \over 2}) + {1 \over 2} (a + b + 1)$. For $c_j, d_j$ integer or half integer, one sees that individually the two ratio of gamma functions in \eqref{5.9} are even polynomial functions of 
$\lambda N_{\rm hJ}'$. Known formulas for the asymptotics of the ratio of gamma functions with shifted arguments gives that
with respect to a large $\lambda N_{\rm hJ}'$
expansion,
the evenness property persists for general $c_j, d_j \in \mathbb R$ fixed relative to 
$\lambda N_{\rm hJ}'$; see the working in
\cite[\S 3.1]{FS26}. Moreover, to leading order the
dependence on $ N_{\rm hJ}'$ 
cancels when combined with the factors coming from
the substitution $\mathbf x \mapsto \mathbf x/(4 (N_{\rm hJ}')^2)$
as required by \eqref{kc1},
as well as the prefactor therein. Recalling too the factor in
\eqref{5.9} independent of $ N_{\rm hJ}'$, and the limit of the
generalised hypergeometric function \eqref{5.9o}, the limit
law \eqref{Trs1} follows.
\end{proof}

\begin{remark}
The limit law \eqref{Trs1} for the hard edge scaled $k$-point correlation is, from \cite[Eq.~(13.47)]{Fo10}, the same as for the hard edge scaling limit $x \mapsto x/(4N)$ of the Laguerre $\beta$ ensemble ($\beta$ even) with one body weight $x^a e^{-\beta x/2} \mathbbm 1_{x > 0}$. 
\end{remark}

 \subsection{The density for $\beta =1$ and $\beta =4$}
 As for the $\beta=1,4$ Laguerre ensembles \cite{FRS26}, it is convenient to introduce the modified $\beta=1$ and 4 Jacobi weight functions
 \begin{equation}\label{H1}
 \begin{cases} x^{{a-1 \over 2}} (1 - x)^{b-1 \over 2} 
 \mathbbm 1_{0 < x < 1}, & \beta = 1, \\
 x^{a+1} (1 - x)^{b+1} \mathbbm 1_{0 < x < 1}, & \beta = 4.
 \end{cases}
 \end{equation}
 The corresponding modifications of \eqref{3.1J} are
 \begin{equation}\label{H2}
\hat{N}_{\rm hJ}' =  \begin{cases}
N  + (a+b-1)/2, & \beta = 1, \\
N + (a+b+1)/4, & \beta = 4.
\end{cases}
\end{equation}
With hard edge scaling variables (recall above \eqref{3.7g}) using this modification of ${N}_{\rm hJ}'$, denote the density
by $\rho_{\beta,N}^{\rm h,J}(y)$. Then, Proposition \ref{prop5.1} implies for $\beta=4$ that
\begin{equation}\label{3.7g4}
\frac{1}{\sqrt{\beta}}\rho_{\beta,N}^{\rm h,J}\Big(\frac{y}{\beta}\Big) = \tilde{r}_{0}^{\rm h,J}(y) +
{1 \over (2 \sqrt{\beta} \hat{N}_{\rm hJ}')^2} \tilde{r}_{1}^{\rm h,J}(y) + 
{1 \over (2 \sqrt{\beta} \hat{N}_{\rm hJ}')^4} \tilde{r}_{2}^{\rm h,J}(y) + \cdots,
\end{equation}
while a $\beta\leftrightarrow4/\beta$ duality \cite{DP12,FRW17},
which leaves $ \hat{N}_{\rm hJ}'$ unchanged up to a factor of 
$-\beta/2$, shows that this form holds true for $\beta=1$. That the coefficients $\tilde{r}_{0}^{\rm h,J}(y), \tilde{r}_{1}^{\rm h,J}(y),\ldots$ are independent
of $\beta$ follows from a similar large $\hat{N}_{\rm hJ}'$ expansion of the fifth order differential operator specifying the Jacobi density for both $\beta=1$ and $4$ \cite{RF19} as made explicit in Appendix B.

\begin{remark}
The density of the Jacobi ensemble exhibits a hard edge at the origin when $a$ is held fixed with respect to $N$, but this hard edge remains when $b$ is allowed to grow proportionally with $N$. Thus, one may wonder if the structure \eqref{3.7g4} is seen in this regime. Inspection of the aforementioned differential equation of \cite{RF19} suggests that this is indeed the case upon redefining $\hat{N}_{\rm hJ}'$ and constraining $b$ to take a certain form: The key mechanism behind the calculations of Appendix B is that the 5th order differential equation characterising the $\beta=1$ and $4$ Jacobi ensemble densities has coefficients that are polynomial in $x,a^2-1,b^2-1,4\beta(\hat{N}_{\rm hJ}')^2$, and these coefficients are even functions of $\hat{N}_{\rm hJ}'$ upon applying the hard edge scaling $x=y/(2\hat{N}_{\rm hJ}')^2$. This phenomenon continues to hold if $\hat{N}_{\rm hJ}'$ is redefined to be as in \eqref{H2}, but with $b$ therein set to zero, and if $b$ is constrained to be proportional to this new value of $\hat{N}_{\rm hJ}'$. A similar observation holds for the JUE, as well.
\end{remark}

In keeping with hard edge universality, the leading order term $\tilde{r}_{0}^{\rm h,J}(y)$ of \eqref{3.7g4} is in fact the same as in the Laguerre case \cite{FRS26}. Hence, extending the set $\{b_1(y),b_2(y),b_3(y)\}$ prescribed by \eqref{4.A} with
\begin{equation}
b_4(y)=\frac{1}{\sqrt{y}}J_a(\sqrt{y}){\rm JI}_a(\sqrt{y}),\quad b_5(y)=J_a'(\sqrt{y}){\rm JI}_a(\sqrt{y}),
\end{equation}
where
\begin{equation}
{\rm JI}_a(y):= \begin{cases}
    \int_y^\infty J_a(t)dt, & \beta = 1, \\
    - \int_0^y J_a(t)dt, & \beta = 4, \end{cases}
\end{equation}
we have for $\beta=1,4$ that
\begin{equation} \label{eq5.6}
\tilde{r}_{0}^{\rm h,J}(y) = \frac{y-a^2}{4}b_1(y)+\frac{1}{4}b_2(y)+\frac{1}{4}b_4(y).
\end{equation}
This, the order in $\hat{N}_{\rm hJ}'$ of the expansion \eqref{3.7g4} to leading and next-to-leading order, and moreover the explicit functional form of the first ($\beta$ independent) density correction $\tilde{r}_{1}^{\rm h,J}(y)$
is confirmed and specified in the  following proposition. 
\begin{prop}\label{P5.2}
With regards to \eqref{3.7g4}, we have
\begin{equation} \label{eq5.7}
\tilde{r}_{1}^{\rm h,J}(y)=\tilde{\alpha}_1^{\rm J}b_1(y)+\tilde{\beta}_1^{\rm J}b_2(y)+\tilde{\gamma}_1^{\rm J}b_3(y)+\tilde{\xi}_1^{\rm J}b_4(y)+\tilde{\eta}_1^{\rm J}b_5(y)
\end{equation}
with low order polynomial coefficients
\begin{multline} \label{eq5.8}
\tilde{\alpha}_1^{\rm J}=\frac{y}{8}(y-a^2-b^2+1),\quad \tilde{\beta}_1^{\rm J}=\frac{y}{12},\quad \tilde{\gamma}_1^{\rm J}=-\frac{y}{12},
\\ \tilde{\xi}_1^{\rm J}=\frac{5y-3a^2-3b^2+3}{24},\quad \tilde{\eta}_1^{\rm J}=\frac{y-a^2-3b^2+1}{24}.
\end{multline}
\end{prop}
\begin{proof}
Following the strategy of \cite[Appendix E]{FRS26}, we read from \cite[Eq.~(6.107) and (6.59)]{Fo10} that the eigenvalue densities of the $\beta=1$ Jacobi orthogonal ensemble (JOE) and $\beta=4$ Jacobi symplectic ensemble (JSE) are respectively given by
\begin{multline*}
\rho_N^{\rm JOE}(x)=\rho_{N-1}^{\rm JUE}(x)+\frac{2N+a+b-2}{2h_{N-2}}x^{\tfrac{a-1}{2}}(1-x)^{\tfrac{b-1}{2}}p_{N-1}(x)\Bigg(\int_0^xt^{\tfrac{a-1}{2}}(1-t)^{\tfrac{b-1}{2}}p_{N-2}(t)\,dt
\\-\frac{2^{a+b+2N-5}\Gamma(\tfrac{N-1}{2})\Gamma(\tfrac{a+N-1}{2})\Gamma(\tfrac{b+N-1}{2})\Gamma(\tfrac{a+b+N-1}{2})}{\pi\Gamma(a+b+2N-3)}\Bigg),
\end{multline*}
\begin{multline*}
\rho_N^{\rm JSE}(x)=\frac{1}{2}\rho_{2N}^{\rm JUE}(x)-\frac{4N+a+b}{4h_{2N-1}}x^{\tfrac{a-1}{2}}(1-x)^{\tfrac{b-1}{2}}p_{2N}(x)\int_0^xt^{\tfrac{a-1}{2}}(1-t)^{\tfrac{b-1}{2}}p_{2N-1}(t)\,dt,
\end{multline*}
where $p_n(x),h_n$ are as in \eqref{3.3J}, \eqref{3.4J} and in the first of these it is required that $N$ be even. Employing the scaling $x=y/(2\hat{N}_{\rm hJ}')^2$ and rewriting all occurrences of $N$ according to \eqref{H2} produces
\begin{equation*}
\rho_{1,N}^{\rm h,J}(y):=\frac{1}{4(\hat{N}_{\rm hJ}')^2}\rho_N^{\rm JOE}\left(\frac{y}{4(\hat{N}_{\rm hJ}')^2}\right),\quad \rho_{4,N}^{\rm h,J}(y):=\frac{1}{4(\hat{N}_{\rm hJ}')^2}\rho_N^{\rm JSE}\left(\frac{y}{4(\hat{N}_{\rm hJ}')^2}\right).
\end{equation*}
Expanding these expressions in large $\hat{N}_{\rm hJ}'$ in the same manner as in the proof of Proposition \ref{P4.2} and with knowledge of $\rho_N^{\rm h,J}(y)$ from \S\ref{S4.4} validates \eqref{eq5.6}, confirms the structure of \eqref{3.7g4} up to the order $1/(\hat{N}_{\rm hJ}')^2$ term, and allows us to read off the coefficients \eqref{eq5.8}.
\end{proof}

We know from \cite{FRS26} that span$\, \{ b_1(y),\dots, b_5(y) \}$ with polynomial coefficients is closed under the action of $\mathcal D_y:= {d \over dy} y$. We also know the explicit form
of the linear combinations for each $\mathcal D_y b_i(y)$
\cite[Eqns.~(4.35) and (5.25)]{FRS26}. This allows us to compute
that
\begin{align}
    \mathcal D_y \tilde{r}_0^{\rm h,J}(y) & = \frac{y}{8}b_1(y) + \frac{1}{8}b_4(y) + \frac{1}{8}b_5(y), \nonumber \\
  \mathcal D_y^2 \tilde{r}_0^{\rm h,J}(y) & =  
   \frac{y}{16}b_1(y) + \frac{y}{16}b_3(y) + \frac{a^2 - y + 1}{16}b_4(y) + \frac{1}{8}b_5(y). 
\end{align}
We can now relate $\tilde{r}_1^{\rm h,J}(y)$ as expressed as
a linear combination with polynomial coefficients of
$\{ b_1(y),\dots, b_5(y) \}$ in Proposition \ref{P5.2} as a differential relation with $\tilde{r}_0^{\rm h,J}(y)$
as analogous to that of Proposition \ref{P4.4} in the case $\beta=2$.

\begin{cor}\label{C5.1}
We have 
\begin{equation}
\tilde{r}^{\rm h,J}_1(y) = {d \over dy} \mathcal P^{\rm *,J}_{1,\beta} \tilde{r}^{\rm h,J}_0(y), \qquad  \mathcal P^{\rm *,J}_{1,\beta}:= -\frac{4}{3} y^2 {d \over d y} + {y \over 3} (y - a^2 - 3b^2 + 1).
\end{equation}
\end{cor}

It is possible to unify the result of Proposition \ref{P4.4} and Corollary \ref{C5.1} by considering the large ${N}_{\rm hJ}'$ hard edge density expansion
\begin{equation}
\rho_{\beta,N}^{\rm h,J}(y) =
\rho_{\beta,\infty,0}^{\rm h}(y) + {1 \over ({N}_{\rm hJ}')^2} \rho_{\beta,\infty,1}^{\rm h}(y) + \cdots,
\end{equation}
for the Jacobi $\beta$ ensemble with weight $x^a(1-x)^b \mathbbm 1_{0 < x < 1}$. Taking into consideration the scale factors in \eqref{3.7g4}, we can verify that the results in question can be written as the one formula
\begin{equation}
\rho_{\beta,\infty,1}^{\rm h}(y) \Big |_{\substack{a = {\beta \over 2}(\hat{a} + 1 - 2/\beta),}{b = {\beta \over 2}(\hat{b} + 1 - 2/\beta)}} =
{d \over d y} Q_{1,\beta}^{*,\rm J}
\rho_{\beta,\infty,0}^{\rm h,J}(y) \Big |_{\substack{a = {\beta \over 2}(\hat{a} + 1 - 2/\beta),}{b = {\beta \over 2}(\hat{b} + 1 - 2/\beta)}},
\end{equation}
with
\begin{equation}\label{Q}
 Q_{1,\beta}^{*,\rm J} :=
 - {1 \over 3 \beta} y^2 {d \over d y} + {y^2 \over 12} +
 {y \over 12} \Big ( - \hat{a}^2 - 3 \hat{b}^2 +
 \Big ( {2 \over \beta} - 1 \Big )^2 \Big ).
\end{equation}

\begin{remark} ${}$ \\
1.~For the Laguerre $\beta$ ensemble with weight $x^{a \beta / 2} e^{-\beta x/2}
\mathbbm 1_{x > 0}$, the analogue of the operator \eqref{Q} has been obtained recently in \cite{FS26}. Denoting by
$ Q_{1,\beta}^{*,\rm L}$ the scaled version of this operator for the Laguerre  $\beta$ ensemble with weight $y^{a}e^{-y} \mathbbm 1_{y > 0}$, we read off from
\cite[Eq.~(3.45)]{FS26} that
\begin{equation}\label{QL}
 Q_{1,\beta}^{*,\rm L} :=
 - {1 \over 6 \beta} y^2 {d \over d y} - {y^2 \over 12} -
 {y \over 24} \Big (  \hat{a}^2 - 
 \Big ( {2 \over \beta} + 1 \Big )^2 \Big ),
\end{equation}
where as in \eqref{Q} we have set $a = {\beta \over 2}(\hat{a} + 1 - 2/\beta)$. \\
2.~The $\beta$ generalisation of \eqref{4.28} is
 \begin{equation}\label{4.28x}
 \mathcal E^{1,\rm h,J}_\beta((0,y);\xi) =
 {\mathcal Q}^{*,\rm J}_{1,\beta} {d \over dy} \mathcal E^{0,\rm h}_\beta((0,y);\xi).
 \end{equation}
 We can test this using the exact result $\beta$-generalising
\eqref{3.23Jb},
 \begin{equation}\label{3.23Jb+}
\mathcal E_{N,\beta}^{\rm h,J}((0,y);\xi,N) \bigg |_{\substack{\xi=1 \\ a=0, \, b = {\beta \over 2}(\hat{b} + 1 - 2/\beta)}} =
\Big ( 1 - y/(2N_{\rm h J}')^2 \Big )^{(\beta/2)((N_{\rm h J}')^2 - (\hat{b}/2)^2)},
\end{equation}
as follows from \cite[Eq.~(1.11)]{FK23}, telling us that with these parameters
\begin{equation}\label{3.23Jbe}
\mathcal E^{0,\rm h}_\beta((0,y);\xi) = e^{-\beta y/8} , \quad 
\mathcal E^{1,\rm h,J}_\beta((0,y);\xi) = e^{-\beta y/8} {\beta \over 32} \Big (
 {y \hat{b}^2 } - {y^2 \over 2} \Big ).
\end{equation}
This is checked to satisfy \eqref{4.28x}.
\end{remark}

\section*{Acknowledgements}
The work of PJF and BJS is supported by a grant from the Australian Research Council, Discovery Project
DP250102552. AAR is supported by Hong Kong RGC grants GRF 16304724 and GRF 17304225.

\section*{Appendix A: The density and spacing distributions for the orthogonal group}
\renewcommand{\thesection}{A} 
\renewcommand{\theHsection}{A} 
\setcounter{equation}{0}
\setcounter{prop}{0}
\setcounter{remark}{0}

Define $\rho^{{\rm h O}^\pm(2N+1)}(Y) = (1/N) K_N^{{\rm  O}^\pm(2N+1)}(Y/N,Y/N)$. According to the discussion and notation from \S \ref{S4.3}, this corresponds to hard edge scaled density for matrices from the classical group
${\rm  O}^\pm(2N+1)$. Use of the explicit functional form
\eqref{3.7e} shows
\begin{equation}\label{A1}
\rho^{{\rm h O}^\pm(2N+1)}(Y) =
{1 \over \pi} \bigg ( 1 \mp {1 \over 2 N}
{\sin 2 Y \over \sin (Y/N)} \bigg ).
\end{equation}
Expanding
\begin{equation}\label{A2}
\rho^{{\rm h O}^\pm(2N+1)}(Y) =
r^{\rm h O^{\pm, o}}_0(Y) + {1 \over N^2} r^{\rm h O^{\pm, o}}_1(Y) + {1 \over N^4} r^{\rm h O^{\pm, o}}_2(Y)
+ \cdots,
\end{equation}
(here the superscript ``o'' in ${\rm O}^{\pm, \rm o}$ indicates underlying odd sized matrices)
we see upon introducing the power series
\begin{equation}
{t \over \sin t} = 1 + \sum_{p=1}^\infty a_{2p} t^{2p},
\quad \Big ( a_2 = {1 \over 6}, \: a_4 = {7 \over 360}, \dots \Big ),
\end{equation}
convergent for $|t| < \pi$, that \eqref{A2} is convergent for $|Y/N| < \pi$ and furthermore
\begin{equation}
r^{\rm h O^{\pm, o}}_j(Y) = \mp \frac{a_{2j}}{\pi} Y^{2j} \bigg ( {\sin 2 Y \over 2 Y} \bigg ).
 \end{equation}
 One consequence of these explicit functional forms is the differential relation
 \begin{equation}\label{A5}
 r^{\rm h O^{\pm, o}}_1(Y) = - {1 \over 24} {d \over d Y} 
 Y^2 {d \over d Y}  r^{\rm h O^{\pm, o}}_0(Y).
 \end{equation}

 Introduce now the probabilities $\{ E^{{\rm O}^\pm(2N+1)}(k;(0,\phi)) \}_{k=0}^N$ for the interval $(0,\phi)$ containing exactly $k$ eigenvalues, and use these to form the generating function
 \begin{equation}\label{A6}
 {\mathcal E}^{{\rm O}^\pm(2N+1)}((0,\phi);\xi) :=
 \sum_{k=0}^\infty (1 - \xi)^k 
 E^{{\rm O}^\pm(2N+1)}(k;(0,\phi)).
 \end{equation}
Analogous to \eqref{A2}, introduce the large $N$ expansion
\begin{equation}\label{A7}
 {\mathcal E}^{{\rm O}^\pm(2N+1)}((0,Y/N);\xi)
 = {\mathcal E}_0^{\rm h O^{\pm, o}}(Y;\xi) + {1 \over N^2}
 {\mathcal E}_1^{\rm h O^{\pm, o}}(Y;\xi)+ {1 \over N^4}
 {\mathcal E}_2^{\rm h O^{\pm, o}}(Y;\xi) + \cdots
 \end{equation}
 Under the assumption of the validity of the hypothesis of Bornemann
\cite{Bo25a} that the $\{{\mathcal E}_j^{\rm h O^{\pm, o}}(Y) \}$ are related to ${\mathcal E}_0^{\rm h O^{\pm, o}}(Y)$ by a differential relation which is independent of $\xi$ (recall the discussion of \S \ref{S4.4}), we have from \eqref{A5} the prediction that
\begin{equation}\label{A8}
{\mathcal E}_1^{\rm h O^{\pm, o}}(Y;\xi) =
- {1 \over 24}
 Y^2 {d^2 \over d Y^2} {\mathcal E}_0^{\rm h O^{\pm, o}}(Y;\xi).
  \end{equation}
  
  A very similar equation has appeared in the recent work \cite{FS26} relating to the quantity analogous to \eqref{A6}
  in the case of random matrices from the classical group $U(N)$. Thus,
  adopting notation consistent with that in \eqref{A6} we set
   \begin{equation}\label{A6U}
 {\mathcal E}^{{\rm U}(N)}((0,\phi);\xi) :=
 \sum_{k=0}^\infty (1 - \xi)^k 
 E^{{\rm U}(N)}(k;(0,\phi)),
 \end{equation}
 and expanding for large $N$ 
 \begin{equation}\label{A7U}
 {\mathcal E}^{{\rm U}(N)}((0,Y/N);\xi)
 = {\mathcal E}_0^{\rm bulk}(Y;\xi) + {1 \over N^2}
 {\mathcal E}_1^{\rm bulk}(Y;\xi)+ {1 \over N^4}
 {\mathcal E}_2^{\rm bulk}(Y;\xi) + \cdots,
 \end{equation}
 we have from \cite[Prop.~2.1]{FS26} that 
 \begin{equation}\label{A9U}
 {\mathcal E}_1^{\rm bulk} (Y;\xi) =
- {1 \over 12}
 Y^2 {d^2 \over d Y^2} {\mathcal E}_0^{\rm bulk}(Y;\xi).
  \end{equation}
  While there is a known identity between generating functions \cite[Eq.~(8.128)]{Fo10}
 \begin{equation}\label{A8U}
 {\mathcal E}^{{\rm U}(2N)}((0,2\phi);\xi) =
 {\mathcal E}^{{\rm O}^+(2N+1)}((0,\phi);\xi)
 {\mathcal E}^{{\rm O}^-(2N+1)}((0,\phi);\xi),
  \end{equation}
  it does not appear to be the case that this together with \eqref{A9U} is sufficient to deduce
  \eqref{A8}.

\begin{remark}\label{RA1} ${}$ \\
    1.~With $\rho^{{\rm h O}^\pm(2N)}(Y) = (1/N_{\rm hJ}') K_N^{{\rm  O}^\pm(2N)}(Y/N_{\rm hJ}',Y/N_{\rm hJ}')$, where $N_{\rm hJ}' = N - 1/2$ (this is consistent with the values noted above
    \eqref{3.7e}, upon noting that in ${\rm O}^-(2N)$ the number of independent eigenvalues is $N-1$),
    for the classical groups ${\rm O}^\pm(2N)$ the analogue of
    \eqref{A1} is
  \begin{equation}\label{A1R}
\rho^{{\rm h O}^\pm(2N)}(Y) =
{1 \over \pi} \bigg ( 1 \pm {1 \over N_{\rm hJ}'}
{\sin 2 Y \over \sin (Y/N_{\rm hJ}')} \bigg ).
\end{equation} 
This follows from the explicit functional forms of $K_N^{{\rm  O}^\pm(2N)}(\theta,\phi)$ in \cite[Prop.~5.5.3]{Fo10}.  Thus analogous to \eqref{2} we can expand
\begin{equation}\label{A14}
\rho^{{\rm h O}^\pm(2N)}(Y) =
r^{\rm h O^{\pm, e}}_0(Y) + {1 \over (N_{\rm hJ}')^2} r^{\rm h O^{\pm, e}}_1(Y) + {1 \over (N_{\rm hJ}')^4} r^{\rm h O^{\pm, e}}_2(Y)
+ \cdots,
\end{equation}
(here the superscript ``e'' in ${\rm O}^{\pm, \rm e}$ indicates underlying even sized matrices). The differential relation
\eqref{A5} remains valid in this case, as then so does \eqref{A8} (subject to the validity of the associated assumption). \\
2.~An expression for $\mathcal E_1^{{\rm hO}^{\pm,o}}(Y;\xi)$ 
in \eqref{A7}, in terms of a second order linear inhomogeneous equation with coefficients involving $\sigma$PIII$'$ transcendents relating to $\mathcal E_0^{{\rm hO}^{\pm,o}}(Y;\xi)$, has been given in \cite[Prop.~5.8]{BFM17}. This can be used to establish \eqref{A8} by following the strategy used to deduce \eqref{A9U} in \cite{FS26}.
 \end{remark}

  \appendix
\section*{Appendix B: Large $\hat{N}_{\rm hJ}'$ expansion of the fifth order linear differential operator for the $\beta = 1,4$
density}
\renewcommand{\thesection}{B} 
\renewcommand{\theHsection}{B} 
\setcounter{equation}{0}
\setcounter{prop}{0}
\setcounter{remark}{0}

We know from \cite[Eq.~(2.25)]{RF19} the explicit form of  two fifth order linear differential operators annihilating the
$\beta =1$ and 4 Jacobi ensemble densities.
Reparametrising for consistency with \eqref{H1}
and expanding for large $\hat{N}_{\rm hJ}'$ we can obtain the
analogue of Proposition \ref{P4.4a}.

\begin{prop}
With $\partial_y=\tfrac{d}{dy}$, define the differential operators
{\small \begin{align*}
\tilde{\mathcal{D}}_0^{\rm h,J}=& 4y^5\partial_y^5+40y^4\partial_y^4+y^3(5y-5a^2+93)\partial_y^3+y^2(19y-22a^2+38)\partial_y^2
\\&+y(y^2+2(4-a^2)y+a^4-16a^2-1)\partial_y+\frac{1}{2}(y^2-(3a^2+1)y+2(a^2-1)^2),
\\ \tilde{\mathcal{D}}_1^{\rm h,J}=&-20y^6\partial_y^5-240y^5\partial_y^4-y^4(20y+5(b^2-4a^2)+772)\partial_y^3-y^3(95y+19b^2-107a^2+691)\partial_y^2
\\&-y^2(3y^2+2(b^2-3a^2+40)y+2b^2(4-a^2)+3a^4-114a^2+111)\partial_y
\\&-\frac{y}{2}(4y^2+(2b^2-11a^2+15)y+7a^4-42a^2-b^2(1+3a^2)+3),
\\ \tilde{\mathcal{D}}_2^{\rm h,J}=&40y^7\partial_y^5+560y^6\partial_y^4+y^5(30y+15(b^2-2a^2)+2251)\partial_y^3+y^4(171y+79b^2-189a^2+2885)\partial_y^2
\\&+y^3(3y^2+2(2b^2-3a^2+100)y+(b^2-2a^2)^2+74b^2-a^4-254a^2+943)\partial_y
\\&+\frac{y^2}{2}(5y^2+(7b^2-13a^2+70)y+b^4-9a^2b^2+13b^2+8a^4-118a^2+78),
\\ \tilde{\mathcal{D}}_3^{\rm h,J}=&-40y^8\partial_y^5-640y^7\partial_y^4-y^6(20y+5(3b^2-4a^2)+3051)\partial_y^3-y^5(133y+101b^2-145a^2+4927)\partial_y^2
\\&-y^4(y^2+2(b^2-a^2+96)y+(b^2-a^2)^2+10(14b^2-23a^2)+2269)\partial_y
\\&-\frac{y^3}{2}(2y^2+5(b^2-a^2+18)y+3(b^2-a^2)^2+5(11b^2-25a^2)+322),
\\ \tilde{\mathcal{D}}_4^{\rm h,J}=&20y^9\partial_y^5+360y^8\partial_y^4+y^7(5y+5(b^2-a^2)+1972)\partial_y^3+y^6(38y+41(b^2-a^2)+3773)\partial_y^2
\\&+y^5(64y+74(b^2-a^2)+2157)\partial_y+\frac{9y^4}{2}(4y+5(b^2-a^2+9)),
\\ \tilde{\mathcal{D}}_5^{\rm h,J}=&-y^5(4y^5\partial_y^5+80y^4\partial_y^4+493y^3\partial_y^3+1078y^2\partial_y^2+719y\partial_y+81).
\end{align*}}
Then, for $\beta=1,4$, $N\ge1$, $a,b>-1$ and $0<y<4\beta(\hat{N}_{\rm hJ}')^2$, the scaled density obeys
\begin{equation}\label{B.1}
\sum_{k=0}^5\frac{1}{(2\sqrt{\beta}\hat{N}_{\rm hJ}')^{2k}}\tilde{\mathcal{D}}_k^{\rm h,J}\rho_{\beta,N}^{\rm h,J}\left(\frac{y}{\beta}\right)=0.
\end{equation}
Inserting the expansion \eqref{3.7g4} in the above and collecting terms of like order in $(\hat{N}_{\rm hJ}')^{-2}$ yields the differential relation
\begin{equation}\label{B.2}
\tilde{\mathcal{D}}_0^{\rm h,J}\tilde{r}_{j}^{\rm h,J}=-\sum_{k=1}^5\tilde{\mathcal{D}}_k^{\rm h,J}\tilde{r}_{j-k}^{\rm h,J},\quad(j=0,1,\ldots),
\end{equation}
where $\tilde{r}_{k}^{\rm h,J}=0$ for $k<0$.
\end{prop}

Equation \eqref{B.2} with $j=1$ is consistent with Proposition \ref{P5.2}. The structure of \eqref{B.1}, with in particular the operators
$\tilde{\mathcal{D}}_k^{\rm h,J}$ independent of $\beta=1,4$, motivates the expansion \eqref{3.7g4} with 
$\{\tilde{r}_{j}^{\rm h,J}(y) \}$ independent
of $\beta=1,4$.
Moreover, the structure of  \eqref{B.2} together with
the span of $\{b_1(y),\ldots,b_5(y)\}$  with polynomial coefficients
being closed under the operation $\mathcal D_y = {d \over d y} y$ \cite{FRS26}, is consistent with the structure seen 
in \eqref{eq5.7} for $j=1$ that all the $\{\tilde{r}_{j}^{\rm h,J}(y) \}$ are members of this span.

 \end{document}